\documentclass[12pt]{article}

\usepackage{amsthm}
\theoremstyle{plain}

\theoremstyle{definition}

\theoremstyle{proposition}

\theoremstyle{lemma}

\theoremstyle{remark}

\usepackage{amsmath}
\usepackage{amssymb}
\usepackage[pdftex]{graphicx}

\makeatletter
 
  \@addtoreset{equation}{section}
 \makeatother

\def\Eq#1{\begin{equation} #1 \end{equation}}
\def\Eqr#1{\begin{eqnarray} #1 \end{eqnarray}}
\def\Eqrsubl#1#2{\begin{subequations}\label{#1}\Eqr{#2}\end{subequations}}

\newcommand{\nn}{\nonumber}
\newcommand{\pd}{\partial}

\newcommand{\bea}{\begin{eqnarray}}
\newcommand{\eea}{\end{eqnarray}}

\def\X5sp{{\rm X}_5}
\def\Y3sp{{\rm Y}_3}
\def\Z3sp{{\rm Z}_3}

\def\mb#1{\mbox{\boldmath $#1$}} 

\begin{document}
\setlength{\oddsidemargin}{0cm}
\setlength{\baselineskip}{7mm}

\begin{titlepage}
\begin{flushright}   \end{flushright} 

~~\\

\vspace*{0cm}
    \begin{Large}
       \begin{center}
         {Path-integrals of perturbative superstrings on curved backgrounds from string geometry theory}
       \end{center}
    \end{Large}
\vspace{1cm}

\begin{center}
              Matsuo S{\sc ato},$^{*}$\footnote
           {
e-mail address : msato@hirosaki-u.ac.jp}
and  Kunihito U{\sc zawa}$^{\dagger}$\footnote
           {
e-mail address : kunihito.uzawa@hirosaki-u.ac.jp}\\
      \vspace{1cm}
       
         {$^{*}$\it Graduate School of Science and Technology, Hirosaki University\\ 
 Bunkyo-cho 3, Hirosaki, Aomori 036-8561, Japan}\\

 {$^{\dagger}$\it
Department of Physics,
School of Science and Technology,
Kwansei Gakuin University, Sanda, Hyogo 669-1337, Japan}\\

 {$^{\dagger}$\it
Research and Education 
Center for Natural Sciences, Keio University, 
Hiyoshi 4-1-1, Yokohama, Kanagawa 223-8521, Japan}\\

\end{center}

\hspace{5cm}

\begin{abstract}
\noindent
String geometry theory is one of the candidates of the non-perturbative formulation of string theory. In this paper, from the string geometry theory,  we derive path-integrals of perturbative superstrings  on all the string backgrounds, $G_{\mu\nu}(x)$ and $B_{\mu\nu}(x)$, by considering fluctuations around the string background configurations, which are parametrized by the string backgrounds.

\end{abstract}

\vfill
\end{titlepage}
\vfil\eject

\setcounter{footnote}{0}

\section{Introduction}\label{intro}
\setcounter{equation}{0}
String geometry theory is one of the candidates of non-perturbative formulation of string theory, based on a path-integral of string manifolds, which are a class of infinite-dimensional manifolds \cite{Sato:2017qhj}. String manifolds are defined by patching open sets of the model space defined by introducing a topology to a set of strings.  One of the remarkable facts concerning string geometry theory is that the path-integral of  the perturbative superstrings in the flat background is derived including the moduli of super Riemann surfaces, by considering fluctuations around the flat background in the theory \cite{Sato:2017qhj, Sato:2019cno,  Sato:2020szq}.

Moreover, configurations of fields in string geometry theory include all configurations of fields in the ten-dimensional supergravities, namely string backgrounds \cite{Honda:2020sbl, Honda:2021rcd}. Especially, it is shown that an infinite number of equations of motion of string geometry theory are consistently truncated to finite numbers of equations of motion of the supergravities. That is, string geometry theory includes string backgrounds not as external fields like the perturbative string theories. Dynamics of string backgrounds are a part of  dynamics of  the fields in the theory. It is natural to expect to derive the path-integral of perturbative strings on the sting backgrounds by considering fluctuations around the corresponding configurations in string geometry theory. 

For each background, one theory is formulated in case of a perturbative string theory, whereas perturbative string theories not only on the flat background but also on non-trivial backgrounds should be derived from a single theory in case of the non-perturbative formulation of string theory. 
Actually, the authors in \cite{Sato:2022owj} derived the path-integrals of perturbative strings on all the string backgrounds $G_{\mu\nu}(x)$, $B_{\mu\nu}(x)$ and $\Phi(x)$ from the bosonic sector of string geometry theory. This paper gives a supersymmetric generalization of this fact.

The organization of the paper is as follows. In section 2, we briefly review the string geometry theory. In section 3,  we set string background configurations parametrized by the string backgrounds $G_{\mu\nu}(x)$ and $B_{\mu\nu}(x)$, and  set the classical part of fluctuations representing strings. In section 4, we consider two-point correlation functions of the quantum part of the fluctuations and derive the path-integrals of the perturbative superstrings on the string backgrounds. In the appendix, we obtain a Green function on the flat superstring manifold.

\vspace{1cm}
%--------section2-------------%
\section{Review of string geometry theory} 
String geometry theory is defined by a partition function
\begin{align}
Z=\int \mathcal{D}\bold{G} \mathcal{D}\Phi  \mathcal{D}\bold{B} \mathcal{D}\bold{A} e^{-S},
\end{align}
where the action is given by
\begin{eqnarray}
S=\int \mathcal{D}\bold{E} \mathcal{D}\bar{\tau} \mathcal{D}\bold{X}_{\hat{D}_T} 
\sqrt{-\bold{G}} \left( e^{-2 \Phi} \left( \mathbf{R}  + 4 \nabla_{\bold{I}} \Phi \nabla^{\bold{I}} \Phi - \frac{1}{2} |\mathbf{H} |^{2}   \right) -\frac{1}{2}\sum_{p=1}^9
|\tilde{\bold{F}}_p|^2     
\right),
\label{action of bos string-geometric model}
\end{eqnarray}
where $\bold{I}=\{d,(\mu \bar{\sigma} \bar{\theta}) \}$, $| \mathbf{H} |^{2}:= \frac{1}{3!} \mathbf{G}^{\bold{I}_{1} \bold{J}_{1}} \mathbf{G}^{\bold{I}_{2} \bold{J}_{2}} \mathbf{G}^{\bold{I}_{3} \bold{J}_{3}} \mathbf{H}_{\bold{I}_{1} \bold{I}_{2} \bold{I}_{3}} \mathbf{H}_{\bold{J}_{1} \bold{J}_{2} \bold{J}_{3}}$, and we use the Einstein notation for the index $\bold{I}$.
The path integral is defined by integrating  a metric $\mathbf{G}_{\bold{I}_{1} \bold{I}_{2}}$, a scalar field $\Phi$, a two-form field $\mathbf{B}_{\bold{I}_{1} \bold{I}_{2}}$ and (k-1)-form fields $\bold{A}_{k-1}$. $\tilde{\bold{F}}_p$ are defined by 
$\sum_{p=1}^9 \tilde{\bold{F}}_p=e^{-\bold{B}_2}\wedge \sum_{k=1}^9 \bold{F}_k$, where $\bold{F}_k$ are field strengths of (k-1)-form fields $\bold{A}_{k-1}$. For example, $\tilde{\bold{F}}_5
=
\bold{F}_5
-\bold{B}_2\wedge 
\bold{F}_3
+\frac{1}{2}\bold{B}_2\wedge \bold{B}_2\wedge \bold{F}_1$. 
They are defined on a {\it Riemannian string manifold}, whose definition is given in \cite{Sato:2017qhj}. 
String manifold is constructed by patching open sets in string model space $E$, whose definition is summarized as follows. 
First, a global time $\bar{\tau}$ is defined canonically and uniquely on a super Riemann surface $\bar{\bold{\Sigma}}$ by the real part of the integral of an Abelian differential uniquely defined on $\bar{\bold{\Sigma}}$ \cite{Krichever:1987a, Krichever:1987b}.
We restrict $\bar{\bold{\Sigma}}$ to a $\bar{\tau}$ constant line and obtain $\bar{\bold{\Sigma}}|_{\bar{\tau}}$. An embedding of $\bar{\bold{\Sigma}}|_{\bar{\tau}}$ to $\mathbb{R}^{d}$ represents a many-body state of superstrings in $\mathbb{R}^{d}$, and is parametrized by coordinates $(\bar{\bold{E}}, \bold{X}_{\hat{D}_{T}}(\bar{\tau}), \bar{\tau})$ where $\bar{\bold{E}}$ is a super vierbein on  $\bar{\bold{\Sigma}}$ and $\bold{X}_{\hat{D}_{T}}(\bar{\tau})$ is a map from  $\bar{\bold{\Sigma}}|_{\bar{\tau}}$ to $\mathbb{R}^{d}$.  $\,\, \bar{}$ represents a representative of the super diffeomorphism and super Weyl transformation on the worldsheet. Giving a super Riemann surface $\bar{\bold{\Sigma}}$ is equivalent to giving a  supervierbein $\bar{\bold{E}}$ up to super diffeomorphism and super Weyl transformations. $\hat{D}_T$ represents all the backgrounds except for the target metric, where T runs IIA, IIB, and I.  IIA GSO projection is attached for T = IIA, and the IIB GSO projection is attached for T = IIB and I as in \cite{Sato:2017qhj}. 
${\boldsymbol X}^{\mu}_{\hat{D}_T}(\bar{\tau}_s)=X^{\mu}+ \bar{\theta}^{\alpha} \psi_{\alpha}^{\mu}+\frac{1}{2} \bar{\theta}^2 F^{\mu}$, where $\mu=0, 1, \cdots d-1$, $\psi_{\alpha}^{\mu}$ is a Majorana fermion and  $F^{\mu}$ is an auxiliary field. We abbreviate $\hat{D}_T$ and $(\bar{\tau}_s)$ of $X^{\mu}$, $\psi_{\alpha}^{\mu}$ and $F^{\mu}$. String model space $E$  is defined by the collection of the string states by considering all the  $\bar{\bold{\Sigma}}$, all the values of $\bar{\tau}$, and all the $\bold{X}_{\hat{D}_{T}}(\bar{\tau})$. How near the two string states is defined by how near the values of $\bar{\tau}$ and how near $\bold{X}_{\hat{D}_{T}}(\bar{\tau})$. 
An $\epsilon$-open neighborhood of $[\bar{{\boldsymbol \Sigma}}, \bold{X}_{\hat{D}_{T}s}(\bar{\tau}_s), \bar{\tau}_s]$ is defined by
\begin{eqnarray}
U([\bar{\bold{E}}, \bold{X}_{\hat{D}_{T}s}(\bar{\tau}_s), \bar{\tau}_s], \epsilon)
:=
\left\{[\bar{\bold{E}}, \bold{X}_{\hat{D}_{T}}(\bar{\tau}), \bar{\tau}] \bigm|
\sqrt{|\bar{\tau}-\bar{\tau}_s|^2
+\| \bold{X}_{\hat{D}_{T}}(\bar{\tau}) -\bold{X}_{\hat{D}_{T} s}(\bar{\tau}_s) \|^2}
<\epsilon   \right\}, \label{SuperNeighbour}
\end{eqnarray}
where $\bar{\bold{E}}$  is a discrete variable in the topology of string geometry.
As a result, $d \bar{\bold{E}}$  cannot be a part of basis that span the cotangent space in (\ref{cotangen}), whereas fields are functionals of $\bar{\bold{E}}$ as in (\ref{LineElement}). The precise definition of the string topology is given in the section 2 in \cite{Sato:2017qhj}.  By this definition, arbitrary two string states on a connected super Riemann surface in $E$ are connected continuously. Thus, there is an one-to-one correspondence between a super Riemann surface in $\mathbb{R}^{d}$ and a curve  parametrized by $\bar{\tau}$ from $\bar{\tau}=-\infty$ to $\bar{\tau}=\infty$ on $E$. That is, curves that represent asymptotic processes on $E$ reproduce the right moduli space of the super Riemann surfaces in $\mathbb{R}^{d}$. Therefore, a string geometry theory possesses all-order information of superstring theory. Indeed, the path integral of perturbative superstrings on the flat spacetime is derived from the string geometry theory as in \cite{Sato:2017qhj, Sato:2020szq}. The consistency of the perturbation theory determines $d=10$ (the critical dimension).
The cotangent space is spanned by 
\begin{eqnarray}
d \bold{X}^{d}_{\hat{D}_{T}} &:=& d \bar{\tau}
\nonumber \\
d \bold{X}^{(\mu \bar{\sigma} \bar{\theta}) }_{\hat{D}_{T}}&:=& d \bold{X}^{\mu}_{\hat{D}_{T}} \left( \bar{\sigma}, \bar{\tau}, \bar{\theta} \right), \label{cotangen}
\end{eqnarray}
where $\mu=0, \dots, d-1$. 
The summation over $(\bar{\sigma}, \bar{\theta})$ is defined by 
$\int d\bar{\sigma}d^2\bar{\theta} \hat{\bold{E}}(\bar{\sigma}, \bar{\tau}, \bar{\theta})$.
$\hat{\bold{E}}(\bar{\sigma}, \bar{\tau}, \bar{\theta})
:=
\frac{1}{\bar{n}}\bar{\bold{E}}(\bar{\sigma}, \bar{\tau}, \bar{\theta})$, where $\bar{n}$ is the lapse function of the two-dimensional metric (See (\ref{ADM}).).
This summation is transformed as a scalar under $\bar{\tau} \mapsto \bar{\tau}'(\bar{\tau}, \bold{X}_{\hat{D}_T}(\bar{\tau}))$ and invariant under a supersymmetry transformation $(\bar{\sigma}, \bar{\theta}) \mapsto (\bar{\sigma}'(\bar{\sigma}, \bar{\theta}), \bar{\theta}'(\bar{\sigma}, \bar{\theta}))$. 
As a result, the action (\ref{action of bos string-geometric model}) is invariant under this $\mathcal{N}=(1,1)$ supersymmetry transformation.
%a supersymmetry transformation,
%\begin{equation}
%(\bar{\sigma}, \bar{\theta}) \mapsto (\bar{\sigma}'(\bar{\sigma}, \bar{\theta}), \bar{\theta}(\bar{\sigma}, \bar{\theta})). \label{supertrans}
%\end{equation}
An explicit form of the line element is given by
\begin{eqnarray}
&&ds^2(\bar{\bold{E}}, \bold{X}_{\hat{D}_{T}}(\bar{\tau}), \bar{\tau}) \nonumber \\
=&&G(\bar{\bold{E}}, \bold{X}_{\hat{D}_{T}}(\bar{\tau}), \bar{\tau})_{dd} (d\bar{\tau})^2 \nonumber \\ 
&&+2 d\bar{\tau} \int d\bar{\sigma} d^2\bar{\theta}  \hat{\bold{E}} \sum_{\mu} G(\bar{\bold{E}}, \bold{X}_{\hat{D}_{T}}(\bar{\tau}), \bar{\tau})_{d \; (\mu \bar{\sigma} \bar{\theta})} d \bold{X}_{\hat{D}_{T}}^{\mu}(\bar{\sigma}, \bar{\tau}, \bar{\theta}) \nonumber \\
&&+\int d\bar{\sigma} d^2\bar{\theta} \hat{\bold{E}}  \int d\bar{\sigma}'  d^2\bar{\theta}' \hat{\bold{E}}'  \sum_{\mu, \mu'} G(\bar{\bold{E}}, \bold{X}_{\hat{D}_{T}}(\bar{\tau}), \bar{\tau})_{ \; (\mu \bar{\sigma} \bar{\theta})  \; (\mu' \bar{\sigma}' \bar{\theta}')} d \bold{X}_{\hat{D}_{T}}^{\mu}(\bar{\sigma}, \bar{\tau}, \bar{\theta}) d \bold{X}_{\hat{D}_{T}}^{\mu'}(\bar{\sigma}', \bar{\tau}, \bar{\theta}'). 
\nonumber \\
&& \label{LineElement}
\end{eqnarray}
The inverse metric $\mathbf{G}^{\bold{I}\bold{J}}(\bar{\bold{E}}, \bold{X}_{\hat{D}_{T}}(\bar{\tau}), \bar{\tau})$\footnote{Like this, the fields $\mathbf{G}_{\bold{I}\bold{J}}$, $\Phi$, $\mathbf{B}_{\bold{L}_1\bold{L}_2}$ and $\bold{A}_{\bold{L}_1 \cdots \bold{L}_{p-1}}$ are functionals of the coordinates $\bar{\bold{E}}$, $\bold{X}_{\hat{D}_{T}}(\bar{\tau})$ and $\bar{\tau}$.} is defined by $\mathbf{G}_{\bold{I}\bold{J}}\mathbf{G}^{\bold{J}\bold{K}}=\mathbf{G}^{\bold{K}\bold{J}}\mathbf{G}_{\bold{J}\bold{I}}=\delta^{\bold{K}}_{\bold{I}}$, where $\delta^{d}_{d}=1$ and $\delta_{\mu \bar{\sigma} \bar{\theta}}^{\mu' \bar{\sigma}' \bar{\theta}'}=\delta_{\mu}^{\mu'} \delta_{\bar{\sigma} \bar{\theta}}^{\bar{\sigma}' \bar{\theta}'}$, where $\delta_{\bar{\sigma} \bar{\theta} }^{\bar{\sigma}' \bar{\theta}'}=\delta_{(\bar{\sigma} \bar{\theta}) (\bar{\sigma}' \bar{\theta}')}=\frac{1}{\hat{\bold{E}}}\delta(\bar{\sigma}-\bar{\sigma}') \delta^2(\bar{\theta}-\bar{\theta}')$. 
In the following, we use $D:=\int d\bar{\sigma} d^2\bar{\theta} \hat{\bold{E}} \delta^{(\mu \bar{\sigma} \bar{\theta})}_{(\mu \bar{\sigma} \bar{\theta})}$, then $\delta^M_M=D+1$. Although $D$ is infinity, we treat $D$ as regularization parameter and will take $D \to \infty$ later.

%%%%%%%%%%%%%%%%%%%%%%%%%%%%%%%%%%%%%%%%%%%%
%%%%%%%%%%%%%%%%%%%%%%%%%%%%%%%%%%%%%%%%%%%%

\section{
Superstring background configurations and fluctuations representing strings}

In this paper, we consider only static configurations, including quantum fluctuations:
\begin{eqnarray}
\partial_d {\bf G_{MN}}&=&0\,,  \nonumber \\
\partial_d {\bf B_{MN}}&=&0\,,  \nonumber \\
\partial_d \mb{\Phi}&=&0\,,  \nn\\
\partial_d {\bf A}_{k-1}&=&0\,. \label{static}
\end{eqnarray}
In this section, we will set classical backgrounds including string backgrounds and consider fluctuations that represent strings around them. Here we fix the charts, where we choose T=IIA, IIB or I.
The Einstein equation of the action (\ref{action of bos string-geometric model}) is given by 
\Eqr{
&& \bar{\bold{R}}_{\bold{M}\bold{N}} -\frac{1}{4} \bar{\bold{H}}_{\bf MAB} \bar{\bold{H}}_{\bold{N}}^{\bf AB}+ 2 \bar{\nabla}_{\bold{M}} \bar{\nabla}_{\bold{N}} \bar{\mb{\Phi}} -\frac{1}{2} \bar{\bold{G}}_{\bold{M}\bold{N}} (\bar{\bold{R}}-4 \bar{\nabla}_{\bold{I}} \bar{\mb{\Phi}} \bar{\nabla}^{\bold{I}} \bar{\mb{\Phi}} +4 \bar{\nabla}_{\bold{I}} \bar{\nabla}^{\bold{I}} {\mb{\Phi}} -\frac{1}{2} |\bar{\bold{H}}|^{2})
\nonumber \\
&& -\frac{1}{2} e^{2 \bar{\Phi}} \sum_{p=1}^9 \Big[ \frac{1}{(p-1)!} \bar{\tilde{\bold{F}}}_{\bold{M} \bold{L}_{1} \cdots \bold{L}_{p-1}} \bar{\tilde{\bold{F}}}_{\bold{N}}^{\bold{L}_{1} \cdots \bold{L}_{p-1}} -\frac{1}{2} \bar{\bold{G}}_{\bold{M}\bold{N}} |\bar{\tilde{\bold{F}}}_{p}|^{2}  \Big]=0,  \label{G}
    \label{Ein}
}
where $\bar{\bold{R}}$, $\bar{\bf R}_{\bf MN}$, 
$\bar{\bf R}^{\bf M}_{\bf NPQ}$, $\bar{\bf \nabla}_{\bf M}$
denote the Ricci scalar, Ricci tensor, curvature 
tensor and covariant derivative constructed from 
the metric $\bar{\bf G}_{\bf MN}$\,. 
We consider a perturbation 
with respect to the metric $\bar{\bf G}_{\bf MN}$:
\Eqr{
\bar{\bf G}_{\bf MN}=\hat{\bf G}_{\bf MN}+\bar{\bf h}_{\bf MN}\,,
   \label{Gh}
}
where $\bar{\bf h}_{\bf MN}$ denotes a fluctuation 
around the 0-th order background $\hat{\bf G}_{\bf MN}$\,. We raise and lower the indices by $\hat{\bf G}_{\bf MN}$ in the following. 
We also consider a perturbation with respect to the 2-form $\bar{\bf B}_{\bf MN}$, (k-1)-form $\bar{\bf A}_{k-1}$\,, and the scalar $\bar{\mb{\Phi}}$ around the 0-th order backgrounds $0$. 

First, we generalize the harmonic gauge when a dilaton couples.   
If we define $\bar{\mb{\psi}}_{\bf MN}$ as
\Eqr{
\bar{\mb{\psi}}_{\bf MN}=\bar{\bf h}_{\bf MN}
-\frac{1}{2}\hat{\bf G}^{\bf IJ}
\bar{\bf h}_{\bf IJ}\hat{\bf G}_{\bf MN}
+\Lambda \hat{\bf G}_{\bf MN}\bar{\mb{\Phi}}\,,
  \label{spsi}
}
the Einstein equations (\ref{Ein}) are written as 
\Eqr{
&& \hat{\bf R}_{\bf MN}-\frac{1}{2}\hat{\bf G}_{\bf MN}\hat{\bf R}
+\frac{1}{2}\left(-\hat{\bf \nabla}_{\bf I}\hat{\bf \nabla}^{\bf I}
\bar{\mb{\psi}}_{\bf MN}
+\hat{\bf R}_{\bf MA}\,{\bar{\mb{\psi}}^{\bf A}}_{\bf N}
+\hat{\bf R}_{\bf NA}\,{\bar{\mb{\psi}}_{\bf M}}^{\bf A}
-2\hat{\bf R}_{\bf MANB}\,\bar{\mb{\psi}}^{\bf AB}\right.\nn\\
&&\left.~~~~~
+\hat{\bf \nabla}_{\bf M}\hat{\bf \nabla}_{\bf A}
{\bar{\mb{\psi}}^{\bf A}}_{\bf N}
+\hat{\bf \nabla}_{\bf N}\hat{\bf \nabla}_{\bf A}
{\bar{\mb{\psi}}_{\bf M}}^{\bf A}
-\hat{\bf \nabla}^{\bf I}\hat{\bf \nabla}^{\bf J}
\bar{\mb{\psi}}_{\bf IJ}\,
\hat{\bf G}_{\bf MN}
+\hat{\bf R}^{\bf IJ}\,
\bar{\mb{\psi}}_{\bf IJ}\,\hat{\bf G}_{\bf MN}
-\hat{\bf R}\,\bar{\mb{\psi}}_{\bf MN}\right)\nn\\
&&~~~~~+\left(2-\Lambda\right)
\hat{\bf \nabla}_{\bf M}\hat{\bf \nabla}_{\bf N}\bar{\mb{\Phi}}
-\left(2-\Lambda\right)
\hat{\bf G}_{\bf MN}\hat{\bf \nabla}_{\bf I}
\hat{\bf \nabla}^{\bf I}\bar{\mb{\Phi}}
=0\,,
   \label{spEin}
}
up to the first order in the fields, 
$\bar{\bf h}_{\bf IJ}$, $\bar{\bf B}_{\bf IJ}$, 
$\bar{\mb{\Phi}}$, and $\bar{\bold{A}}_{k-1}$. 
$\hat{\bf R}$, $\hat{\bf R}_{\bf MN}$, 
${\hat{\bf R}^{\bf M}}_{\bf ~~NPQ}$, $\hat{\bf \nabla}_{\bf M}$
denote the Ricci scalar, Ricci tensor, curvature 
tensor and covariant derivative constructed from 
the metric $\hat{\bf G}_{\bf MN}$\,. 
We set $\Lambda=2$ so that the Einstein equation becomes only 
for $\bar{\mb{\psi}}_{\bf MN}$.
$\bar{\bf h}_{\bf MN}$ is inversely expressed as 
\Eqr{
\bar{\bf h}_{\bf MN}
=\bar{\mb{\psi}}_{\bf MN}
+\frac{1}{D-1}\left(-\hat{\bf G}^{\bf PQ}
\,\bar{\mb{\psi}}_{\bf PQ}
+4\bar{\mb{\Phi}}\right)
\hat{\bf G}_{\bf MN}. \label{inverse}
}
We impose a generalization of the harmonic gauge: 
\Eqr{
\hat{\bf \nabla}^{\bf M}\bar{\mb{\psi}}_{\bf MN}=0,
    \label{nsh}
}
which reduces to the ordinary harmonic gauge if the dilaton is zero.
Then, the Einstein equation (\ref{spEin}) becomes 
\Eqr{
&&\hat{\bf R}_{\bf MN}-\frac{1}{2}\hat{\bf G}_{\bf MN}\hat{\bf R}
+\frac{1}{2}\left(-\hat{\bf \nabla}_{\bf I}\hat{\bf \nabla}^{\bf I}
\bar{\mb{\psi}}_{\bf MN}
+\hat{\bf R}_{\bf MA}\,{\bar{\mb{\psi}}^{\bf A}}_{\bf N}
+\hat{\bf R}_{\bf NA}\,{\bar{\mb{\psi}}_{\bf M}}^{\bf A}
-2\hat{\bf R}_{\bf MANB}\,\bar{\mb{\psi}}^{\bf AB}\right.\nn\\
&&\left.~~~~~
+\hat{\bf R}^{\bf IJ}\,
\bar{\mb{\psi}}_{\bf IJ}\,\hat{\bf G}_{\bf MN}
-\hat{\bf R}\,\bar{\mb{\psi}}_{\bf MN}\right)
=0\,.
   \label{nspEin2-2}
}
Next, we set the 0-th order background $\hat{\bf G}_{\bf MN}$ as a flat background:
\Eqr{
\label{metric}
&&\hat{\bf G}_{\bf MN}=a_{\bf M}\,\eta_{\bf MN}\,, \label{flatmetric}
}
where $a_d=1$ and $a_{(\mu\bar{\sigma}\bar{\theta})}=\frac{\bar{e}^3(\bar{\sigma})}
{\sqrt{\bar{h}(\bar{\sigma})}}$. Then, the gauge fixing condition (\ref{nsh}) becomes
\Eq{
\int d\bar{\sigma}\,d^2\bar{\theta}\,\hat{\mb{E}}\,%(\sigma, \theta)\,
\pd^{(\mu\bar{\sigma}\bar{\theta})}
\bar{\mb{\psi}}_{(\mu\bar{\sigma}\bar{\theta}) \bf M}
%g_{(\mu\sigma\theta)(\mu'\sigma'\theta')}
=0\,,
\label{GaugeMetric}
}
the Einstein equation (\ref{nspEin2-2})  becomes Laplace equation,
\Eqr{
\int d\bar{\sigma}\,d^2\bar{\theta}
\hat{\mb{E}}\,
\pd_{(\mu\bar{\sigma}\bar{\theta})}
\pd^{(\mu\bar{\sigma}\bar{\theta})}
\,\bar{\mb{\psi}}_{\bf MN}=0\, ,
   \label{nspEin3}
}
and the components of (\ref{inverse}) read
\Eqr{
\bar{\bf h}_{dd}&=&\frac{D-2}{D-1}\,\bar{\mb{\psi}}_{dd}
+\frac{1}{D-1}\int d\bar{\sigma}''\,d^2\bar{\theta}''
\hat{\mb{E}}''\,
{\bar{\mb{\psi}}^{(\mu''\bar{\sigma}''\bar{\theta}'')}}_
{(\mu''\bar{\sigma}''\bar{\theta}'')}
-\frac{4}{D-1}\,\bar{\mb{\Phi}}\,,  \nonumber\\
\bar{\bf h}_{d(\mu\bar{\sigma}\bar{\theta})}&=&
\bar{\mb{\psi}}_{d(\mu\bar{\sigma}\bar{\theta})}\,,  \nonumber \\
\bar{\bf h}_{(\mu\bar{\sigma}\bar{\theta})
(\mu'\bar{\sigma}'\bar{\theta}')}&=&
\bar{\mb{\psi}}_{(\mu\bar{\sigma}\bar{\theta})
(\mu'\bar{\sigma}'\bar{\theta}')}
+\frac{\bar{e}^3}
{\sqrt{\bar{h}}}\,
\delta_{(\mu\bar{\sigma}\bar{\theta})
(\mu'\bar{\sigma}'\bar{\theta}')}
\left(\frac{1}{D-1}\,\bar{\mb{\psi}}_{dd}
\right.\nn\\
&&\left.~~~~~
-\frac{1}{D-1}\int d\bar{\sigma}''\,
\,d^2\bar{\theta}''\,\hat{\mb{E}}''\,
{\bar{\mb{\psi}}^{(\mu''\bar{\sigma}''\bar{\theta}'')}}
_{(\mu''\bar{\sigma}''\bar{\theta}'')}
+\frac{4}{D-1}\bar{\mb{\Phi}}\right)\,. \label{compo}
}

Next, the equation of motion of the scalar of the action (\ref{action of bos string-geometric model}) 
\Eqr{
\bar{\bold{R}}-4 \bar{\bf \nabla}_{\bold{M}} 
\bar{\mb{\Phi}} \partial^{\bold{M}} \bar{\mb{\Phi}} 
+4 \bar{\bf \nabla}_{\bold{M}} \bar{\bf \nabla}^{\bold{M}} 
\bar{{\mb{\Phi}}} -\frac{1}{2} |\bar{\bold{H}}|^{2}=0,   
   \label{deq}
}
is written as 
\Eqr{
{\bf \hat{R}+\hat{\nabla}^M\hat{\nabla}^N 
\bar{h}_{MN}
-\hat{\nabla}^M\hat{\nabla}_M {\bar{h}^N}_N
+4\hat{G}^{MN}\hat{\nabla}_M
\hat{\nabla}_N\bar{\mb{\Phi}}}=0\,,
   \label{deq1}
}
up to the first order in the fields, 
$\bar{\bf h}_{\bf IJ}$, $\bar{\bf B}_{\bf IJ}$, 
$\bar{\bf A}_{k-1}$ and $\bar{\mb{\Phi}}$. 
Furthermore, this can be written as
\Eqr{
&&\int d\bar{\sigma}\,d^2\bar{\theta}\,
\hat{\mb{E}}\,
\pd_{(\mu\bar{\sigma}\bar{\theta})}
\pd^{(\mu\bar{\sigma}\bar{\theta})}\bar{\mb{\Phi}}
+\frac{1}{4}\int d\bar{\sigma}\,d^2\bar{\theta}
\,\hat{\mb{E}}\,
\pd_{(\mu\bar{\sigma}\bar{\theta})}
\pd^{(\mu\bar{\sigma}\bar{\theta})}\,\bar{\mb{\psi}}_{dd}\nn\\
&&~~~
-\frac{1}{4}\int d\bar{\sigma}\,d^2\bar{\theta}\,
\,\hat{\mb{E}}\,
\pd_{(\mu\bar{\sigma}\bar{\theta})}
\pd^{(\mu\bar{\sigma}\bar{\theta})}\,
\int d\bar{\sigma}'\,d^2\bar{\theta}'\,
\hat{\mb{E}}'\,
{\bar{\mb{\psi}}^{(\mu'\bar{\sigma}'\bar{\theta}')}}
_{(\mu'\bar{\sigma}'\bar{\theta}')}
=0\,,
}
around the flat 0-th order background (\ref{flatmetric}) under the static condition (\ref{static}) in the generalized harmonic gauge (\ref{nsh}). This becomes Laplace equation,
\Eq{
\int d\bar{\sigma}\,d^2\bar{\theta}\,
\hat{\mb{E}}\,
\pd_{(\mu\bar{\sigma}\bar{\theta})}
\pd^{(\mu\bar{\sigma}\bar{\theta})}\bar{\mb{\Phi}}
=0\,, \label{eomscalar}
}
if the metric satisfies the Einstein equation (\ref{nspEin3}).

On the other hand, the equation of motion of B-field
and the (k-1)-form fields 
\Eqr{
%{\bf \bar{\nabla}_M}(e^{-2\bar{\phi}} {\bf \bar{H}^{MNP}})=0
 &&\bar{\bf \nabla}_{\bold{M}} ( e^{-2\bar{\Phi}} \bar{\bold{H}}^{\bf MNP}) \nonumber \\
&&+\sum_{p=3}^9 \sum_{n=0}^{[\frac{p-3}{2}]} \frac{1}{2^{n+1} \cdot (p-2)!}
 \bar{\bold{F}}_{ \bold{I}_{1} \cdots \bold{I}_{p-2-2n} } 
\bar{\bold{B}}_{\bold{J}_{1} \bold{K}_{1} } \cdots \bar{\bold{B}}_{\bold{J}_{n} \bold{K}_{n} } 
 \bar{\tilde{\bold{F}}}^{ \bold{I}_{1} \cdots \bold{I}_{p-2-2n} \bold{J}_{1} \bold{K}_{1}  \cdots \bold{J}_{n} \bold{K}_{n}  \bold{L}_{1}\bold{L}_{2} }=0,  \nonumber \\
  \label{B} \\
&& \bar{\bf \nabla}_{\bold{I}} \bar{\tilde{\bold{F}}}^{\bold{I} \bold{L}_{1} \cdots \bold{L}_{p-1}} + \left( \frac{1}{2} \right)^{n} \hat{\bf \nabla}_{\bold{I}} \Big[  \bar{\bold{B}}_{\bold{J}_{1} \bold{K}_{1} } \cdots \bar{\bold{B}}_{\bold{J}_{n} \bold{K}_{n} } \bar{\tilde{\bold{F}}}^{\bold{J}_{1}\bold{K}_{1} \cdots \bold{J}_{n}\bold{K}_{n} \bold{I} \bold{L}_{1} \cdots \bold{L}_{p-1} }   \Big]=0, \label{A}
}
are written as 
\Eqr{
&&\hat{\bf \nabla}_{\bf M} \bar{\bf H}^{\bf MNP}=0\,,
  \label{B1}\\
&&\hat{\bf \nabla}_{\bold{I}} \bar{\tilde{\bold{F}}}^{\bold{I} 
\bold{L}_{1} \cdots \bold{L}_{p-1}}=0\,, 
}
up to the first order in the fields, $\bar{\bf h}_{\bf IJ}$, $\bar{\bf B}_{\bf IJ}$, $\bar{\bf A}_{k-1}$ and $\bar{\mb{\Phi}}$. Furthermore, the equation (\ref{B1}) becomes Laplace equation
\Eqr{
\int d\bar{\sigma}\,d^2\bar{\theta}\,
\hat{\mb{E}}\,
\pd_{(\mu\bar{\sigma}\bar{\theta})}\pd^{(\mu\bar{\sigma}\bar{\theta})}
\,\bar{\bf B}_{\bf MN}=0\,. \label{eomB}
}
around the flat 0-th order background (\ref{flatmetric}) under the static condition (\ref{static}) in Lorentz gauge,
\begin{equation}
\hat{\bf \nabla}_{\bf M} \bar{\bf B}^{\bf MN}=0, 
\end{equation}
which is equivalent to
\begin{equation}
\partial_{(\mu\bar{\sigma}\bar{\theta})} 
\bar{\bf B} ^{(\mu\bar{\sigma}\bar{\theta}) {\bf N}}=0 \label{BLorentz}
\end{equation}

We consider classical solutions corresponding to the string background configurations where R-R fields are set zero:
\begin{eqnarray}
\bar{\mb{\psi}}_{dd}&=&0  \\
\bar{\mb{\psi}}_{d(\mu\bar{\sigma}\bar{\theta})}&=&0  \label{psidmu}\\
\bar{\mb{h}}_{(\mu\bar{\sigma}\bar{\theta})(\mu'\bar{\sigma}'\bar{\theta}')}&=&\frac{\bar{e}^3}{\sqrt{\bar{h}}}\,g_{\mu\nu}({\bf X}_{\hat{D}_{T}}(\bar{\sigma}, \bar{\theta}))\delta_{\bar{\sigma}\bar{\sigma}'}\,\delta_{\bar{\theta}\bar{\theta}'},  \label{psimunu}\\
\bar{\bf B}_{d(\mu\bar{\sigma}\bar{\theta})}&=&0 \label{Bdmu} \\
\bar{\bf B}_{(\mu\bar{\sigma}\bar{\theta})(\mu'\bar{\sigma}'\bar{\theta}')}&=&\frac{\bar{e}^3}{\sqrt{\bar{h}}}\,
B_{\mu\nu}({\bf X}_{\hat{D}_{T}}(\bar{\sigma}, \bar{\theta}))
\delta_{\bar{\sigma}\bar{\sigma}'}\delta_{\bar{\theta}\bar{\theta}'},  \label{Bmunu}\\
\bar{\mb{\Phi}}&=& \int d \bar{\sigma}\,d^2\bar{\theta}\, \hat{{\bf E}} \,{\Phi}({\bf X}_{\hat{D}_{T}}(\bar{\sigma}, \bar{\theta})), \label{Phi}\\
\bar{\bf A}_{k-1}&=&0\,, \label{A=0}
\end{eqnarray}
where $g_{\mu\nu}(x)$, $B_{\mu\nu}(x)$ and  $\Phi(x)$   satisfy Laplace equations,
\begin{eqnarray}
\partial_{\rho} \partial^{\rho} g_{\mu\nu}(x)&=&0 \nonumber \\
\partial_{\rho} \partial^{\rho} B_{\mu\nu}(x)&=&0 \nonumber \\
\partial_{\rho} \partial^{\rho} \Phi(x)&=&0,  \label{Laplacesugra}
\end{eqnarray}
and gauge fixing conditions,
\begin{eqnarray}
\partial^{\mu} \psi_{\mu\nu}(x)&=&0 \nonumber \\
\partial^{\mu} B_{\mu\nu}(x)&=&0, \label{gaugefixing10d}
\end{eqnarray}
where 
\Eqr{
\psi_{\mu \nu}=g_{\mu \nu}
-\frac{1}{2}\delta^{\alpha \beta}
g_{\alpha \beta}\delta_{\mu \nu}
+2 \delta_{\mu \nu}\Phi\,,
}
which imply (\ref{GaugeMetric}), (\ref{nspEin3}), (\ref{eomscalar}), (\ref{eomB}) and (\ref{BLorentz}).
Indeed,  these are equivalent to
\begin{eqnarray}
\bar{\bf G}_{dd}&=&-1  \\
\bar{\bf G}_{d(\mu\bar{\sigma}\bar{\theta})}&=&0 \\
\bar{\bf G}_{(\mu\bar{\sigma}\bar{\theta})(\mu'\bar{\sigma}'\bar{\theta}')}&=&\frac{\bar{e}^3}{\sqrt{\bar{h}}}\,
G_{\mu\nu}({\bf X}_{\hat{D}_{T}}(\bar{\sigma}, \bar{\theta}))
\delta_{\bar{\sigma}\bar{\sigma}'}\,\delta_{\bar{\theta}\bar{\theta}'}, \\
\bar{\bf B}_{d(\mu\bar{\sigma}\bar{\theta})}&=&0  \\
\bar{\bf B}_{(\mu\bar{\sigma}\bar{\theta})(\mu'\bar{\sigma}'\bar{\theta}')}&=&\frac{\bar{e}^3}{\sqrt{\bar{h}}}\,
B_{\mu\nu}({\bf X}_{\hat{D}_{T}}(\bar{\sigma}, \bar{\theta}))
\delta_{\bar{\sigma}\bar{\sigma}'}\,\delta_{\bar{\theta}\bar{\theta}'}, \\
\bar{\mb{\Phi}}&=& \int d \bar{\sigma}\,d^2\bar{\theta}\,\hat{\bf E}\,\Phi({\bf X}_{\hat{D}_{T}}(\bar{\sigma}, \bar{\theta})),\\
\bar{\bf A}_{k-1}&=&0\,,
\end{eqnarray}
where
\Eq{
G_{\mu\nu}({\bf X}_{\hat{D}_{T}})=\delta_{\mu\nu}+g_{\mu\nu}({\bf X}_{\hat{D}_{T}}) \,.
}
These are the string background configurations themselves \cite{Honda:2020sbl, Honda:2021rcd}.
(\ref{Laplacesugra}) implies that
$\bar{\bf G}_{\bf MN}$, $\bar{\bf B}_{\bf MN}$, $\bar{\bf A}_{k-1}$ and $\bar{\mb{\Phi}}$ satisfy their equations of motion in string geometry theory\footnote{Under (\ref{eomscalar}),  
(\ref{nspEin3}) is equivalent to $
\int d\bar{\sigma}\,d^2\bar{\theta}
\hat{\mb{E}}\,
\pd_{(\mu\bar{\sigma}\bar{\theta})}
\pd^{(\mu\bar{\sigma}\bar{\theta})}
\,\bar{\mb{h}}_{\bf MN}=0$, because (\ref{inverse}).}
(\ref{nspEin3}), (\ref{eomscalar}) and (\ref{eomB}), and 
$G_{\mu \nu}$, $B_{\mu \nu}$, and $\Phi$ also satisfy their equations of motion in the supergravity.
Therefore, these string background configurations  in string geometry theory 
 represent perturbative string vacua parametrized by the on-shell fields in the supergravity as string backgrounds.

Next, we consider fluctuations around these vacua. The scalar fluctuation $\mb{\psi}_{dd}$ represents the degrees of freedom of  perturbative strings in the case of the flat background as in \cite{Sato:2017qhj, Sato:2019cno, Sato:2020szq}. Thus, we also consider the scalar fluctuation $\mb{\psi}_{dd}$ around the perturbative vacua. We set the classical part of $\mb{\psi}_{dd}$ as
\begin{equation}
\bar{\mb{\psi}}_{dd} ({\bf X}_{\hat{D}_{T}}) =\int {\cal D}{\bf X'}_{\hat{D}_{T}} G({\bf X}_{\hat{D}_{T}} ; {\bf X'}_{\hat{D}_{T}})
\omega ({\bf X'}_{\hat{D}_{T}})\,,
 \label{scalarclassicalfluctuation}
\end{equation}
where we will choose a particular function $\omega ({\bf X}_{\hat{D}_{T}})$ later 
%we have chosen $\omega$:
and 
$G({\bf X}_{\hat{D}_{T}} ; {\bf X'}_{\hat{D}_{T}})$ is a Green function on the flat superstring manifold given by 
\begin{equation}
G({\bf X}_{\hat{D}_{T}} ; {\bf X'}_{\hat{D}_{T}})= \mathcal{N} \left[\int d\bar{\sigma}' d^2\bar{\theta}'
\frac{\bar{e}^{'2}}{\bar{\bf E}'}\left({\bf X}_{\hat{D}_{T}}^\mu(\bar{\sigma}', \bar{\theta}')
-{\bf X'}_{\hat{D}_{T}}^\mu(\bar{\sigma}', \bar{\theta}')\right)^2
\right]^{\frac{2-D}{2}}\,, \label{GreenFunc}
\end{equation}
which satisfies 
\begin{equation}
\int d\bar{\sigma}d^2\bar{\theta}
\,\bar{\bf{E}}\,
\frac{1}{\bar{e}}\frac{\pd}{\pd {\bf X}_{\hat{D}_{T}}^\mu(\bar{\sigma}, \bar{\theta})}
\frac{1}{\bar{e}}\frac{\pd}{\pd {\bf X}_{\hat{D}_{T}\,\mu}(\bar{\sigma}, \bar{\theta})}
G({\bf X}_{\hat{D}_{T}} ; {\bf X'}_{\hat{D}_{T}})=\delta({\bf X}_{\hat{D}_{T}}-{\bf X'}_{\hat{D}_{T}}), \label{GreenEq}
\end{equation}
where $\mathcal{N}$ is a normalizing constant.
A derivation is given in the appendix.  
As a result, $\bar{\mb{\psi}}_{dd}$ is not on-shell but satisfies
\Eqr{
\int d\bar{\sigma}d^2\bar{\theta}
\,\bar{{\bf E}}\,
\frac{1}{\bar{e}}\frac{\pd}{\pd {\bf X}_{\hat{D}_{T}}^\mu(\bar{\sigma}, \bar{\theta})}
\frac{1}{\bar{e}}\frac{\pd}{\pd {\bf X}_{\hat{D}_{T}\,\mu}(\bar{\sigma}, \bar{\theta})}
\,\bar{\mb{\psi}}_{dd}({\bf X}_{\hat{D}_{T}})
=\omega ({\bf X}_{\hat{D}_{T}})\,.
 \label{NoOnshell}
}

Furthermore, we consider the quantum part of $\mb{\psi}_{dd}$,
\begin{equation}
\tilde{\mb{\psi}}_{dd}=\frac{D-1}{D-2}\tilde{\phi},
\end{equation}
where $\frac{D-1}{D-2}$ is introduced for later convenience. 
Totally, 
\Eqr{
{\bf G}_{\bf MN}=
\hat{\bf G}_{\bf MN}+\bar{\bf h}_{\bf MN}+\tilde{\bf G}_{\bf MN}\,, 
\label{G1}
}
where $\hat{\bf G}_{\bf MN}$  is  given by (\ref{flatmetric}), ${\bf \bar{h}_{MN}}$  is given by (\ref{compo}) with (\ref{scalarclassicalfluctuation}), (\ref{psidmu}), (\ref{psimunu}) and (\ref{Phi}), and ${\bf \tilde{G}_{MN}}$ is given by
\Eqr{
&&\hspace{-1cm}
\tilde{\bf G}_{dd}=\tilde{\phi}\,,~~~~
\tilde{\bf G}_{d(\mu\bar{\sigma}\bar{\theta})}=0\,,~~~~
\tilde{\bf G}_{(\mu\bar{\sigma}\bar{\theta})(\mu'\bar{\sigma}'\bar{\theta}')}
=\frac{1}{D-2}
\frac{\bar{e}^3}{\sqrt{\bar{h}}}\,
\tilde{\phi}\,\delta_{(\mu\bar{\sigma}\bar{\theta})(\mu'\bar{\sigma}'\bar{\theta}')}\,.
  \label{gh}
}

%%%%%%%%%%%%%%%%%%%%%%%%%%%%%%%%%%%%%%%%%%%%
%%%%%%%%%%%%%%%%%%%%%%%%%%%%%%%%%%%%%%%%%%%%
%%%%%%%%%%%%%%%%%%%%%%%%%%%%%%%%%%%%%%%%%%%%
%%%%%%%%%%%%%%%%%%%%%%%%%%%%%%%%%%%%%%%%%%%%
%%%%%%%%%%%%%%%%%%%%%%%%%%%%%%%%%%%%%%%%%%%%
%
%======================================%
%<<<<<<<<<<<<< SECTION B >>>>>>>>>>>>>>%
%======================================%
%
\section{Deriving the path-integrals of the perturbative superstrings on curved backgrounds
}
In this section, we will derive the path integrals of the perturbative superstrings in any order from the tree-level two-point correlation functions of the quantum scalar fluctuations of the metric. 
In order to obtain a propagator, we add a gauge fixing term corresponding to 
(\ref{nsh}) into the action (\ref{action of bos string-geometric model}) and obtain
\Eqr{
S&=&
\int \mathcal{D}\bar{\tau} \mathcal{D}\bold{E} 
\mathcal{D}\bold{X}_{\hat{D}_{T}} 
\sqrt{-\bold{G}} \left[e^{-2 \Phi} \left( \mathbf{R}  
+ 4 \nabla_{\bold{I}} {\mb{\Phi}} \nabla^{\bold{I}} {\mb{\Phi}} 
- \frac{1}{2} |\mathbf{H} |^{2}   \right) 
-\frac{1}{2}\sum_{p=1}^9|\tilde{\bold{F}}_p|^2   
\right.\nn\\
&&\left.
-\frac{1}{2}\left\{{\bf \bar{\nabla}^N\left(\tilde{G}_{MN}-
\frac{1}{2} \bar{G}^{IJ} \tilde{G}_{IJ}\bar{G}_{MN}
+2\bar{G}_{MN}\,\bar{\mb{\Phi}}\right)}\right\}^2
\right],
  \label{s-action}
}
where we abbreviate the  Faddeev-Popov ghost term because it does not contribute to the tree-level two-point correlation functions of the metrics. 
By substituting Eqs.~(\ref{G1}), (\ref{Bdmu}),  
(\ref{Bmunu}),  (\ref{Phi})  and (\ref{A=0}) that do not necessarily satisfy the equations of motion (\ref{Laplacesugra}) into (\ref{s-action}),  
this is expressed as 
\Eqr{
S&=&\int 
 {\cal D}\bar{\tau}\, {\cal D}{{\bf E}}\,{\cal D}{\bf X}_{\hat{D}_{T}}
\,
\left(c_0+c_1\,\tilde{\phi}
+\tilde{\phi}\,c_2\,\tilde{\phi}\right.\nn\\
&&\left.
+\tilde{\phi}\, \int d\bar{\sigma}\,d^2\bar{\theta}\,\hat{\bf E}\,
\int d\bar{\sigma}'\,d^2\bar{\theta}'\,\hat{\bf E}'
c^{(\mu\bar{\sigma}\bar{\theta})(\mu'\bar{\sigma}'\bar{\theta}')}
\,\pd_{(\mu\bar{\sigma}\bar{\theta})}
\pd_{(\mu'\bar{\sigma}'\bar{\theta}')}\tilde{\phi}
\right)\,, \label{fixedaction}
}
where 
\Eqrsubl{c}{
&&\hspace{-1cm}c_0=
-\frac{1}{D-1}\int d\bar{\sigma}\,d^2\bar{\theta}\,\hat{\bf E}\,
\pd_{(\mu\bar{\sigma}\bar{\theta})}\pd^{(\mu\bar{\sigma}\bar{\theta})}\,
\bar{\mb{\psi}}_{dd}
-\frac{4D}{D-1}\int d\bar{\sigma}\,d^2\bar{\theta}\,\hat{\bf E}\,
\pd_{(\mu\bar{\sigma}\bar{\theta})}
\pd^{(\mu\bar{\sigma}\bar{\theta})}\bar{\mb{\Phi}}\nn\\
&&~~~~~+\frac{1}{D-1}\int d\bar{\sigma}\,d^2\bar{\theta}\,\hat{\bf E}\,
\pd_{(\mu\bar{\sigma}\bar{\theta})}\pd^{(\mu\bar{\sigma}\bar{\theta})}
\int d\bar{\sigma}'\,d^2\bar{\theta}'\,\hat{\bf E}'\,
{\bar{\mb{\psi}}_{(\mu'\bar{\sigma}'\bar{\theta}')}}^
{(\mu'\bar{\sigma}'\bar{\theta}')}\,,\\
&&\hspace{-1cm}c_1=
\frac{1}{2}\int d\bar{\sigma}\,d^2\bar{\theta}\,
\hat{\bf E}\,
\pd_{(\mu\bar{\sigma}\bar{\theta})}
\pd^{(\mu\bar{\sigma}\bar{\theta})}\,\bar{\mb{\psi}}_{dd}\nn\\
&&
+\frac{1}{2(D-2)}\int d\bar{\sigma}\,d^2\bar{\theta}\,\hat{\bf E}\,
\pd_{(\mu\bar{\sigma}\bar{\theta})}\pd^{(\mu\bar{\sigma}\bar{\theta})}
\int d\bar{\sigma}'\,d^2\bar{\theta}'\,\hat{\bf E}'\,
{\bar{\mb{\psi}}_{(\mu'\bar{\sigma}'\bar{\theta}')}}^{(\mu'\bar{\sigma}'\bar{\theta}')}\,,\\
&&\hspace{-1cm}c_2=
\frac{1}{4}\int d\bar{\sigma}\,d^2\bar{\theta}\,\hat{\bf E}\,
\pd_{(\mu\bar{\sigma}\bar{\theta})}
\pd^{(\mu\bar{\sigma}\bar{\theta})}\,\bar{\mb{\psi}}_{dd}\nn\\
&&
-\frac{1}{4(D-2)^2}\int d\bar{\sigma}\,d^2\bar{\theta}\,\hat{\bf E}\,
\pd_{(\mu\bar{\sigma}\bar{\theta})}\pd^{(\mu\bar{\sigma}\bar{\theta})}\,
\int d\bar{\sigma}'\,d^2\bar{\theta}\,\hat{\bf E}\,
{\bar{\mb{\psi}}_{(\mu'\bar{\sigma}'\bar{\sigma}')}}^{(\mu'\bar{\sigma}'\bar{\sigma}')}\,,\\
&&\hspace{-1cm}
c^{(\mu\bar{\sigma}\bar{\theta})(\mu'\bar{\sigma}'\bar{\theta}')}=
\left[
\frac{D-1}{4(D-2)}
+\frac{1}{2}
\,\bar{\mb{\psi}}_{dd}
+\frac{1}{2(D-2)}\int d\bar{\sigma}''\,d^2\bar{\theta}''\,\hat{\bf E}''\,
{\bar{\mb{\psi}}_{(\mu''\bar{\sigma}''\bar{\theta}'')}}^{(\mu''\bar{\sigma}''\bar{\theta}'')}
\right.\nn\\
&&\left.~~~~~
-\frac{2}{D-2}
\bar{\mb{\Phi}}
\right]
\delta^{(\mu\bar{\sigma}\bar{\theta})(\mu'\bar{\sigma}'\bar{\theta}')}
-\frac{D-1}{4(D-2)}
\bar{\mb{\psi}}^{(\mu\bar{\sigma}\bar{\theta})(\mu'\bar{\sigma}'\bar{\theta}')}\,,
}
up to the first order in the classical fields and the second order in $\tilde{\phi}$.
In $D\rightarrow\infty$, (\ref{fixedaction}) is
\Eqr{
\hspace{-0.5cm}S&=&\int 
 {\cal D}\bar{\tau}\, {\cal D}{\bf E}\,{\cal D}{\bf X}_{\hat{D}_{T}}\,
\left[-4\int d\bar{\sigma}\,d^2\bar{\theta}\,
\hat{\bf E}\,\pd_{(\mu\bar{\sigma}\bar{\theta})}
\pd^{(\mu\bar{\sigma}\bar{\theta})}\bar{\mb{\Phi}}
+\frac{1}{2}\,\int d\bar{\sigma}\,d^2\bar{\theta}\,\hat{\bf E}\,
\pd_{(\mu\bar{\sigma}\bar{\theta})}
\pd^{(\mu\bar{\sigma}\bar{\theta})}\bar{\mb{\psi}}_{dd}
\,\tilde{\phi}
\right.\nn\\
&&\hspace{-0.7cm}
+\,\tilde{\phi}\,\frac{1}{4}\int d\bar{\sigma}\,
d^2\bar{\theta}\,\hat{\bf E}\,
\pd_{(\mu\bar{\sigma}\bar{\theta})}
\pd^{(\mu\bar{\sigma}\bar{\theta})}\bar{\mb{\psi}}_{dd}
\,\tilde{\phi}
+\tilde{\phi}\,
\left(\frac{1}{4}+\frac{1}{2}\bar{\mb{\psi}}_{dd}\right)
\int d\bar{\sigma}\,d^2\bar{\theta}\,\hat{\bf E}\,
\,\pd_{(\mu\bar{\sigma}\bar{\theta})}
\pd^{(\mu\bar{\sigma}\bar{\theta})}\tilde{\phi}
\nn\\
&&\left.\hspace{-0.7cm}
-\frac{1}{4}\tilde{\phi}\,\int d\bar{\sigma}\,
d^2\bar{\theta}\,\hat{\bf E}\,
\int d\bar{\sigma}'\,d^2\bar{\theta}'\,\hat{\bf E}'\,
\bar{\mb{\psi}}^{(\mu\bar{\sigma}\bar{\theta})(\mu'\bar{\sigma}'\bar{\theta}')}
\,\pd_{(\mu\bar{\sigma}\bar{\theta})}
\pd_{(\mu'\bar{\sigma}'\bar{\theta}')}\tilde{\phi}
\right].
  \label{action2}
}
By shifting the field $\tilde{\phi}$ as $\tilde{\phi}=\tilde{\phi}'
-\frac{2}{3}$\,, the first order term in $\tilde{\phi}'$ vanishes as
\Eqr{
\hspace{-0.5cm}S&=&\int 
 {\cal D}\bar{\tau}\, {\cal D}{\bf E}\,{\cal D}{\bf X}_{\hat{D}_{T}}\,
\biggl[\tilde{\phi}'\,\frac{1}{4}\int d\bar{\sigma}\,
d^2\bar{\theta}\,\hat{\bf E}\,
\pd_{(\mu\bar{\sigma}\bar{\theta})}
\pd^{(\mu\bar{\sigma}\bar{\theta})}\bar{\mb{\psi}}_{dd}
\,\tilde{\phi}'
\nonumber \\
&&
+\tilde{\phi}'\,
\left(\frac{1}{4}+\frac{1}{2}\bar{\mb{\psi}}_{dd}+\frac{1}{8}\hat{\bf G}^{\bf IJ}
\bar{\bf h}_{\bf IJ} -\frac{1}{2}\bar{\mb{\Phi}}\right)
\int d\bar{\sigma}\,d^2\bar{\theta}\,\hat{\bf E}\,
\,\pd_{(\mu\bar{\sigma}\bar{\theta})}
\pd^{(\mu\bar{\sigma}\bar{\theta})}\tilde{\phi}'
\nn\\
&&\left.\hspace{-0.7cm}
-\frac{1}{4}\tilde{\phi}'\,\int d\bar{\sigma}\,
d^2\bar{\theta}\,\hat{\bf E}\,
\int d\bar{\sigma}'\,d^2\bar{\theta}'\,\hat{\bf E}'\,
\bar{\mb{h}}^{(\mu\bar{\sigma}\bar{\theta})(\mu'\bar{\sigma}'\bar{\theta}')}
\,\pd_{(\mu\bar{\sigma}\bar{\theta})}
\pd_{(\mu'\bar{\sigma}'\bar{\theta}')}\tilde{\phi}'
\right].
  \label{action-1st-2}
}
where surface terms are dropped and the gauge fixing condition in (\ref{gaugefixing10d}) and a relation (\ref{spsi}) are applied.
By normalizing the leading part of the kinetic term as 
$\tilde{\phi}'=-2(1-\bar{\mb{\psi}}_{dd}-\frac{1}{4}\hat{\bf G}^{\bf IJ}
\bar{\bf h}_{\bf IJ}+\bar{\mb{\Phi}})\tilde{\phi}''$\,, 
we have 
\Eqr{
S&=&\int 
 {\cal D}\bar{\tau}\, {\cal D}{\bf E}\,{\cal D}{\bf X}_{\hat{D}_{T}}
\left[\int d\bar{\sigma}\,d^2\bar{\theta}\,\hat{\bf E}\,
\,\pd_{(\mu\bar{\sigma}\bar{\theta})}\pd^{(\mu\bar{\sigma}\bar{\theta})}
\bar{\mb{\psi}}_{dd}\,
\left(\tilde{\phi}''\right)^2
+
\tilde{\phi}''\,\int d\bar{\sigma}\,d^2\bar{\theta}\,\hat{\bf E}\,
\pd_{(\mu\bar{\sigma}\bar{\theta})}\pd^{(\mu\bar{\sigma}\bar{\theta})}\tilde{\phi}''
\right.
\nn\\
&&~~~\left.
-\tilde{\phi}''\int d\bar{\sigma}\,d^2\bar{\theta}\,\hat{\bf E}\,\int d\bar{\sigma}'d^2\bar{\theta}'\,\hat{\bf E}'\,
\,\bar{\mb{h}}_{(\mu\bar{\sigma}\bar{\theta})(\mu'\bar{\sigma}'\bar{\theta}')}\,
\pd^{(\mu\bar{\sigma}\bar{\theta})}\pd^{(\mu'\bar{\sigma}'\bar{\theta}')}
\tilde{\phi}''
\right].
  \label{action-1st-3}
}

In the following, we consider only the case that  the quantum fluctuation is local as
\Eqr{
&&\tilde{\phi}''=\int d\bar{\sigma}\,d^2\bar{\theta}\,
f\left({\bf X}_{\hat{D}_{T}}(\bar{\sigma}, \bar{\theta})\right)\,,
}
and obtain component representation of (\ref{action-1st-3}). 
Under the super diffeomorphism transformation of  
$\bar{\sigma}$ and $\bar{\theta}$, 
$\int  d \bar{\sigma} \bar{e}$
and
$\int  d \bar{\sigma} d^2\bar{\theta} \hat{\bold{E}}$
are invariant, then 
$\frac{1}{\bar{e}}\delta(\bar{\sigma}-\bar{\sigma}')$
and
$\frac{1}{\hat{\bold{E}}}\delta(\bar{\sigma}-\bar{\sigma}') 
\delta^2(\bar{\theta}-\bar{\theta}')$
are scalars, thus
$\frac{\bar{e}}{\hat{\bold{E}}}\delta^2(\bar{\theta}-\bar{\theta}')$
is a scalar. Hence we have 
\begin{eqnarray}
\frac{\pd}{\pd {X}^{\mu}(\bar{\sigma})}
&=&\int d\sigma' d^2\bar{\theta}'\,\hat{\mb{E}}(\bar{\sigma}', \bar{\theta}')\,
\frac{\pd {\bf X}_{\hat{D}_{T}}^{\nu}(\bar{\sigma}', \bar{\theta}')} 
{\pd {X}^{\mu}(\bar{\sigma})}
\frac{\pd}{\pd {\bf X}_{\hat{D}_{T}}^{\nu}(\bar{\sigma}', \bar{\theta}')} \nonumber \\
&=&\int d\sigma' d^2\bar{\theta}'\,\hat{\mb{E}}(\bar{\sigma}', \bar{\theta}')\,
\frac{1}{\bar{e}}\delta^{\nu}_{\mu}\delta(\bar{\sigma}-\bar{\sigma}')
\frac{\pd}{\pd {\bf X}_{\hat{D}_{T}}^{\nu}(\bar{\sigma}', \bar{\theta}')} \nonumber \\
&=&\int d^2\bar{\theta}\,\frac{\hat{\mb{E}}(\bar{\sigma}, \bar{\theta})}
{\bar{e}(\bar{\sigma})}\,
\frac{\pd}{\pd {\bf X}_{\hat{D}_{T}}^{\mu}(\bar{\sigma}, \bar{\theta})}, 
\end{eqnarray}
\begin{eqnarray}
\frac{\pd}{\pd {X}^{\mu}(\bar{\sigma})}
\frac{\pd}{\pd {X}^{\nu}(\bar{\sigma})}\tilde{\phi}''
&=&
\int d^2\bar{\theta}'\,\frac{\hat{\mb{E}}(\bar{\sigma}, \bar{\theta}')}
{\bar{e}(\bar{\sigma})}\,
\int d^2\bar{\theta}\,\frac{\hat{\mb{E}}(\bar{\sigma}, \bar{\theta})}
{\bar{e}(\bar{\sigma})}\,
\frac{\pd}{\pd {\bf X}_{\hat{D}_{T}}^{\mu}(\bar{\sigma}, \bar{\theta}')} 
\frac{\pd}{\pd {\bf X}_{\hat{D}_{T}}^{\nu}(\bar{\sigma}, \bar{\theta})} 
\tilde{\phi}'' \nonumber \\
&=&
\int d^2\bar{\theta}'\,\frac{\hat{\mb{E}}(\bar{\sigma}, \bar{\theta}')}
{\bar{e}(\bar{\sigma})}\,
\int d^2\bar{\theta}\,\frac{\hat{\mb{E}}(\bar{\sigma}, \bar{\theta})}
{\bar{e}(\bar{\sigma})}
\frac{\bar{e}(\bar{\sigma})}{\hat{\mb{E}}(\bar{\sigma}, \bar{\theta})}
\delta(\bar{\theta}-\bar{\theta}')
\frac{\pd}{\pd {\bf X}_{\hat{D}_{T}}^{\mu}(\bar{\sigma}, \bar{\theta})} 
\frac{\pd}{\pd {\bf X}_{\hat{D}_{T}}^{\nu}(\bar{\sigma}, \bar{\theta})} 
\tilde{\phi}'' \nonumber \\
&=&\int d^2\bar{\theta}\,\frac{\hat{\mb{E}}(\bar{\sigma}, \bar{\theta})}
{\bar{e}(\bar{\sigma})}\,
\frac{\pd}{\pd {\bf X}_{\hat{D}_{T}}^{\mu}(\bar{\sigma}, \bar{\theta})}
\frac{\pd}{\pd {\bf X}_{\hat{D}_{T}}^{\nu}(\bar{\sigma}, \bar{\theta})}
\tilde{\phi}'', \nonumber 
\end{eqnarray}
\begin{eqnarray}
\frac{\pd}{\pd {\psi}^{\mu}(\bar{\sigma})}
&=&\int d\sigma' d^2\bar{\theta}'\,\hat{\mb{E}}(\bar{\sigma}', \bar{\theta}')\,
\frac{\pd {\bf X}_{\hat{D}_{T}}^{\nu}(\bar{\sigma}', \bar{\theta}')} 
{\pd {\psi}^{\mu}(\bar{\sigma})}
\frac{\pd}{\pd {\bf X}_{\hat{D}_{T}}^{\nu}(\bar{\sigma}', \bar{\theta}')} \nonumber \\
&=&\int d\sigma' d^2\bar{\theta}'\,\hat{\mb{E}}(\bar{\sigma}', \bar{\theta}')\,
\frac{1}{\bar{e}}\delta^{\nu}_{\mu}\delta(\bar{\sigma}-\bar{\sigma}')
\bar{\theta}'
\frac{\pd}{\pd {\bf X}_{\hat{D}_{T}}^{\nu}(\bar{\sigma}', \bar{\theta}')} \nonumber \\
&=&\int d^2\bar{\theta}\,\frac{\hat{\mb{E}}(\bar{\sigma}, \bar{\theta})}
{\bar{e}(\bar{\sigma})}\,
\bar{\theta}
\frac{\pd}{\pd {\bf X}_{\hat{D}_{T}}^{\mu}(\bar{\sigma}, \bar{\theta})}, 
\end{eqnarray}
\begin{eqnarray}
\frac{\pd}{\pd {X}^{\mu}(\bar{\sigma})}
\frac{\pd}{\pd {\psi}^{\nu}(\bar{\sigma})}\tilde{\phi}''
&=&
\int d^2\bar{\theta}'\,\frac{\hat{\mb{E}}(\bar{\sigma}, \bar{\theta}')}
{\bar{e}(\bar{\sigma})}\,
\int d^2\bar{\theta}\,\frac{\hat{\mb{E}}(\bar{\sigma}, \bar{\theta})}
{\bar{e}(\bar{\sigma})}\,
\frac{\pd}{\pd {\bf X}_{\hat{D}_{T}}^{\mu}(\bar{\sigma}, \bar{\theta}')} 
\bar{\theta}
\frac{\pd}{\pd {\bf X}_{\hat{D}_{T}}^{\nu}(\bar{\sigma}, \bar{\theta})} 
\tilde{\phi}'' \nonumber \\
&=&
\int d^2\bar{\theta}'\,\frac{\hat{\mb{E}}(\bar{\sigma}, \bar{\theta}')}
{\bar{e}(\bar{\sigma})}\,
\int d^2\bar{\theta}\,\bar{\theta} \frac{\hat{\mb{E}}(\bar{\sigma}, \bar{\theta})}
{\bar{e}(\bar{\sigma})}
\frac{\bar{e}(\bar{\sigma})}{\hat{\mb{E}}(\bar{\sigma}, \bar{\theta})}
\delta(\bar{\theta}-\bar{\theta}')
\frac{\pd}{\pd {\bf X}_{\hat{D}_{T}}^{\mu}(\bar{\sigma}, \bar{\theta})} 
\frac{\pd}{\pd {\bf X}_{\hat{D}_{T}}^{\nu}(\bar{\sigma}, \bar{\theta})} 
\tilde{\phi}'' \nonumber \\
&=&\int d^2\bar{\theta}\, \bar{\theta} \frac{\hat{\mb{E}}(\bar{\sigma}, \bar{\theta})}
{\bar{e}(\bar{\sigma})}\,
\frac{\pd}{\pd {\bf X}_{\hat{D}_{T}}^{\mu}(\bar{\sigma}, \bar{\theta})}
\frac{\pd}{\pd {\bf X}_{\hat{D}_{T}}^{\nu}(\bar{\sigma}, \bar{\theta})}
\tilde{\phi}'', \nonumber 
\end{eqnarray}
\begin{eqnarray}
\frac{\pd}{\pd {X}^{\mu}(\bar{\sigma})}
\frac{\pd}{\pd {F}^{\nu}(\bar{\sigma})}\tilde{\phi}''
&=&
\int d^2\bar{\theta}'\,\frac{\hat{\mb{E}}(\bar{\sigma}, \bar{\theta}')}
{\bar{e}(\bar{\sigma})}\,
\int d^2\bar{\theta}\,\frac{\hat{\mb{E}}(\bar{\sigma}, \bar{\theta})}
{\bar{e}(\bar{\sigma})}\,
\frac{\pd}{\pd {\bf X}_{\hat{D}_{T}}^{\mu}(\bar{\sigma}, \bar{\theta}')} 
\frac{1}{2}\bar{\theta}^2
\frac{\pd}{\pd {\bf X}_{\hat{D}_{T}}^{\nu}(\bar{\sigma}, \bar{\theta})} 
\tilde{\phi}'' \nonumber \\
&=&
\int d^2\bar{\theta}'\,\frac{\hat{\mb{E}}(\bar{\sigma}, \bar{\theta}')}
{\bar{e}(\bar{\sigma})}\,
\int d^2\bar{\theta}\,\frac{1}{2}\bar{\theta}^2 \frac{\hat{\mb{E}}(\bar{\sigma}, \bar{\theta})}
{\bar{e}(\bar{\sigma})}
\frac{\bar{e}(\bar{\sigma})}{\hat{\mb{E}}(\bar{\sigma}, \bar{\theta})}
\delta(\bar{\theta}-\bar{\theta}')
\frac{\pd}{\pd {\bf X}_{\hat{D}_{T}}^{\mu}(\bar{\sigma}, \bar{\theta})} 
\frac{\pd}{\pd {\bf X}_{\hat{D}_{T}}^{\nu}(\bar{\sigma}, \bar{\theta})} 
\tilde{\phi}'' \nonumber \\
&=&\int d^2\bar{\theta}\, \frac{1}{2}\bar{\theta}^2 \frac{\hat{\mb{E}}(\bar{\sigma}, \bar{\theta})}
{\bar{e}(\bar{\sigma})}\,
\frac{\pd}{\pd {\bf X}_{\hat{D}_{T}}^{\mu}(\bar{\sigma}, \bar{\theta})}
\frac{\pd}{\pd {\bf X}_{\hat{D}_{T}}^{\nu}(\bar{\sigma}, \bar{\theta})}
\tilde{\phi}'', \nonumber 
\end{eqnarray}
and
\Eqr{
g_{\mu\nu}({\bf X}_{\hat{D}_{T}}(\bar{\sigma}, \bar{\theta}))
=g_{\mu\nu}(X)+\pd_\rho g_{\mu\nu}(X)\,\bar{\theta}\,\psi^\rho\
+\frac{1}{2}\bar{\theta}^2(\pd_\rho g_{\mu\nu}(X)F^{\rho}
+\frac{1}{2}\pd_\rho \pd_{\rho'} g_{\mu\nu}(X) \psi^{\rho} \psi^{\rho'}).
}
Collecting the above results, the action 
(\ref{action-1st-3}) is expressed as 
\Eqr{
&&
S=\int 
 {\cal D}\bar{\tau}\, {\cal D}h\,{\cal D}{X}\,
\tilde{\phi}'' \Biggl(
\int d\bar{\sigma}\,\sqrt{\bar{h}}%\,\frac{1}{e^2}
\left(%\tilde{\phi}''\,
\left(\delta_{\mu\nu}-g_{\mu\nu}\right)
\,\frac{1}{\bar{e}^2}\,
\frac{\pd}{\pd {X}_{\mu}}
\frac{\pd}{\pd {X}_{\nu}}
\tilde{\phi}''\right.
-
\pd_\rho g_{\mu\nu}(X)\,\psi^\rho
\,\frac{1}{\bar{e}^2}\,
\frac{\pd}{\pd {X}_{\mu}}
\frac{\pd}{\pd \psi_{\nu}}\tilde{\phi}''
\nn\\
&&\left.
\hspace{1cm}
-(\pd_\rho g_{\mu\nu}(X)F^{\rho}
+\frac{1}{2}\pd_\rho \pd_{\rho'} g_{\mu\nu}(X) \psi^{\rho} \psi^{\rho'})
\,\frac{1}{\bar{e}^2}\,
\frac{\pd}{\pd {X}_{\mu}}
\frac{\pd}{\pd F_{\nu}}\tilde{\phi}''\right)
+\omega \tilde{\phi}''\, \Biggr).
}
This can be written as
\Eqr{
S=-2
\int {\cal D}\bar{\tau}\,{\cal D}h
\,{\cal D}X\,\tilde{\phi}''\,
H\left(-i\frac{1}{\bar{e}}\frac{\pd}{\pd X}\,,
\frac{\pd}{\pd \psi}\,,
\frac{\pd}{\pd F}\,,
\bold{X}_{\hat{D}_{T}}\,,\bar{\bold{E}}\right)
\tilde{\phi}''\,,
}
where
\Eqr{
&&H\left(-i\frac{1}{\bar{e}}\frac{\pd}{\pd X}\,,
\frac{\pd}{\pd \psi}\,,
\frac{\pd}{\pd F}\,,
\bold{X}_{\hat{D}_{T}}\,,\bar{\bold{E}}\right) \nn\\
&=&\frac{1}{2}\int d\bar{\sigma}\,\sqrt{\bar{h}}
\left(\delta^{\mu\nu}-g^{\mu\nu}\right)
\left({-i\frac{1}{\bar{e}}\frac{\pd}{\pd X^\mu}}\right)\,\left({-i\frac{1}{\bar{e}}\frac{\pd}{\pd X^\nu}}\right)
\nn\\
&&+\int d\bar{\sigma}\, 
n^{\bar{\sigma}} \pd_{\bar{\sigma}} X^\mu \,
{\bar{e}}\,\left({-i\frac{1}{\bar{e}}\frac{\pd}{\pd X^\mu}}\right)
+\int d\bar{\sigma}  
\,i\,\frac{\sqrt{\bar{h}}}{\bar{e}}\,
\pd_{\bar{\sigma}} X^\nu {B_{\nu}}^\mu\,
\left({-i\frac{1}{\bar{e}}\frac{\pd}{\pd X^\mu}}\right)
\nn\\
&&
-\int d\bar{\sigma}\,\frac{\bar{h}}{\bar{e}}\,
\bar{\psi}^\nu \frac{i}{2}\left(
{\frac{1}{\sqrt{\bar{h}}}\Gamma_\nu^{\mu\rho}}(X)\gamma^0\,(\frac{\pd}{\pd \psi^\rho})+{{H_\nu}^{\mu}_{\,\,\,\rho}}(X)\,
\gamma_5 \gamma^0\,\psi^\rho\right)
\,\left({-i\frac{1}{\bar{e}}\frac{\pd}{\pd X^\mu}}\right)\nn\\
&&+\frac{i}{2}\int d\bar{\sigma}\,i\,\frac{\bar{h}}
{\bar{e}}\,
\left(-2\,\bar{\chi}_a\,\gamma^0\,\gamma^a\,\psi^\mu
\right)
\,\left({-i\frac{1}{\bar{e}}\frac{\pd}{\pd X^\mu}}\right)
\nn\\
&&+\int d\bar{\sigma}\,\sqrt{\bar{h}}\frac{i}{2\bar{e}}(\pd_\rho g_{\mu\nu}(X)F^{\rho}
+\frac{1}{2}\pd_\rho \pd_{\rho'} g_{\mu\nu}(X) \psi^{\rho} \psi^{\rho'})
\,
(-i\frac{1}{\bar{e}}\frac{\pd}{\pd {X}_{\mu}})
(\frac{\pd}{\pd F_{\nu}})-\frac{1}{2}
\omega\,. 
}
Here 
$\Gamma^\rho_{\mu\nu}=\frac{1}{2}G^{\rho\lambda}
\left(\pd_\mu G_{\nu\lambda}+\pd_\nu G_{\mu\lambda}
-\pd_\lambda G_{\mu\nu}\right)$\,,
$\bar{\psi}^{\mu}=\psi^{\mu} \gamma^0$\,, 
$\gamma^0=\left(
\begin{array}{cc}
0 & -i \\
i & 0
\end{array}
\right)
$\,, $\gamma_5=\left(
\begin{array}{cc}
-1 & 0 \\
0 & 1
\end{array}
\right)$\,, 
and we have added terms 
\Eqr{
&&\hspace{-0.5cm}
0=-2
\int {\cal D}\bar{\tau}\,{\cal D}h\,{\cal D}X\,\tilde{\phi}''\,
\left[-i\int d\bar{\sigma} \bar{n}^{\bar{\sigma}} \pd_{\bar{\sigma}} X^\mu
\frac{\pd}{\pd X^\mu}
+\int d\bar{\sigma} \frac{\sqrt{\bar{h}}}{\bar{e}^2}\,
\pd_{\bar{\sigma}} X^\nu 
{B_{\nu}}^\mu\frac{\pd}{\pd X^\mu}
\right.
\nn\\
&&+\frac{i}{4}\int d\bar{\sigma}\,\frac{\bar{h}}
{\bar{e}^2}\,i\,\bar{\psi}^\mu \partial_{\nu}{{H_\mu}^\nu}_\rho(X)\,
\gamma_5\,\gamma^0\,\psi^\rho
\nonumber \\
&&\left.
+\frac{i}{2}\int d\bar{\sigma}\,\frac{\bar{h}}
{\bar{e}^2}\,i\,\bar{\psi}^\mu {{H_\mu}^\nu}_\rho(X)\,
\gamma_5\,\gamma^0\,\psi^\rho\frac{\pd}{\pd X^\nu}
+\frac{i}{2}\int d\bar{\sigma}\,\frac{\bar{h}}{\bar{e}^2}\,
\left(-2\bar{\chi}_a\,\gamma^0\,\gamma^a\,\psi^\mu\,
\frac{\pd}{\pd X^\mu}\right)
\right]
\tilde{\phi}''\,,
\nn\\
\label{insert}}
which is true because of the gauge fixing condition (\ref{gaugefixing10d}).

The propagator for $\tilde{\phi}$ defined by
\begin{equation}
\Delta_F(\bar{\bold{E}}, \bold{X}_{\hat{D}_{T}}(\bar{\tau}); \; \bar{\bold{E}},'  \bold{X}_{\hat{D}_{T}}'(\bar{\tau}'))=<\tilde{\phi} (\bar{\bold{E}}, \bold{X}_{\hat{D}_{T}}(\bar{\tau}))
\tilde{\phi}(\bar{\bold{E}},'  \bold{X}_{\hat{D}_{T}}'(\bar{\tau}'))>\,,
\end{equation}
satisfies
\begin{eqnarray}
&&H\left(-i\frac{1}{\bar{e}}\frac{\pd}{\pd X(\bar{\tau})}\,,
\frac{\pd}{\pd \psi(\bar{\tau})}\,,
\frac{\pd}{\pd F(\bar{\tau})}\,,
\bold{X}_{\hat{D}_{T}}(\bar{\tau})\,,\bar{\bold{E}}\right)
\Delta_F(\bar{\bold{E}}, \bold{X}_{\hat{D}_{T}}(\bar{\tau}); \; \bar{\bold{E}},'  \bold{X}_{\hat{D}_{T}}'(\bar{\tau}')) \nn \\
&=&
\delta(\bar{\bold{E}}-\bar{\bold{E}}') \delta(\bold{X}_{\hat{D}_{T}}(\bar{\tau})-\bold{X}'(\bar{\tau}')). \label{delta}
\end{eqnarray}

In order to obtain a Schwinger representation of the propagator, we use the operator formalism $(\hat{\bar{\bold{E}}},  \hat{\bold{X}}_{\hat{D}_{T}}(\bar{\tau}))$ of the first quantization. 
%An operator ordering can be taken appropriately because the Hamiltonian is second order at most. 
The eigen state for $(\hat{\bar{\bold{E}}},  \hat{X}( \bar{\tau}))$ is given by $|\bar{\bold{E}}, X(\bar{\tau})>$. The conjugate momentum is written as $(\hat{\bold{p}}_{\bar{\bold{E}}}, \hat{p}_{X})$. There is no conjugate momentum for the auxiliary field $F^{\mu}$, whereas the Majorana fermion $\psi^{\mu}_{\alpha}$ is self-conjugate and satisfy $\{ \hat{\psi}_{\alpha}^{\mu}(\bar{\sigma}), \hat{\psi}_{\beta}^{\nu}(\bar{\sigma}') \}= \frac{1}{\sqrt{\bar{h}}}\delta_{\alpha \beta} G^{\mu\nu}(X) \delta(\bar{\sigma}-\bar{\sigma}')$.  By defining creation and annihilation operators for $\psi^{\mu}_{\alpha}$ as $\hat{\psi}^{\mu \dagger}:= \frac{1}{\sqrt{2}}(\hat{\psi}^{\mu}_1-i\hat{\psi}^{\mu}_2)$ and $\hat{\psi}^{\mu}:= \frac{1}{\sqrt{2}}(\hat{\psi}^{\mu}_1+i\hat{\psi}^{\mu}_2)$,
one obtains an algebra $\{ \hat{\psi}^{\mu}(\bar{\sigma}), \hat{\psi}^{\nu \dagger}(\bar{\sigma}') \}= \frac{1}{\sqrt{\bar{h}}} G^{\mu\nu}(X) \delta(\bar{\sigma}-\bar{\sigma}')$, $\{ \hat{\psi}^{\mu}(\bar{\sigma}), \hat{\psi}^{\nu}(\bar{\sigma}') \}= 0$, and $\{ \hat{\psi}^{\mu \dagger}(\bar{\sigma}), \hat{\psi}^{\nu \dagger}(\bar{\sigma}') \}=0$. The vacuum $|0>$ for this algebra is defined by $\hat{\psi}^{\mu}(\bar{\sigma}) |0>=0$. The eigen state $|\psi>$, which satisfies $\hat{\psi}^{\mu}(\bar{\sigma}) |\psi>= \psi^{\mu}(\bar{\sigma}) |\psi>$, is given by 
$
e^{-\psi \cdot \hat{\psi}^{\dagger}} |0>
:=
e^{- \int d\bar{\sigma} \sqrt{\bar{h}}G_{\mu\nu}(X) \psi^{\mu}(\bar{\sigma}) \hat{\psi}^{\nu \dagger}(\bar{\sigma})} |0>$. Then, the inner product is given by $<\psi | \psi'>=e^{\psi^{\dagger} \cdot \psi'}$, whereas the completeness relation is
$\int \mathcal{D}\psi^{\dagger} \mathcal{D}\psi |\psi>e^{-\psi^{\dagger} \cdot \psi}<\psi|=1$.

Since \eqref{delta} means that $\Delta_F$ is an inverse of $H$, $\Delta_F$ can be expressed by a matrix element of the operator $\hat{H}^{-1}$ as
\begin{equation}
\Delta_F(\bar{\bold{E}}, \bold{X}_{\hat{D}_{T}}(\bar{\tau}); \; \bar{\bold{E}},'  \bold{X}_{\hat{D}_{T}}'(\bar{\tau}'))
=
<\bar{\bold{E}}, \bold{X}_{\hat{D}_{T}}(\bar{\tau})| H^{-1}(\hat{p}_{X}(\bar{\tau}), \sqrt{\hat{\bar{h}}}\hat{G}_{\mu\nu}\,\hat{\psi}^\nu(\bar{\tau}), 0, \hat{\bold{X}}_{\hat{D}_{T}}(\bar{\tau}), \hat{\bar{\bold{E}}}) |\bar{\bold{E}},' \bold{X}_{\hat{D}_{T}}'(\bar{\tau}') >,
\label{InverseH}
\end{equation}
where we use the fact that the states do not depend on $F^{\mu}$ because it is not an independent valuable and written in terms of the other fields $X^{\mu}$ and $\psi^{\mu}$. 

On the other hand,
\begin{eqnarray}
\hat{H}^{-1}=  i \int _0^{\infty} dT e^{-iT\hat{H}}, \label{IntegralFormula}
\end{eqnarray}
because
\begin{equation}
\lim_{\epsilon \to 0+} \int _0^{\infty} dT e^{-T(i\hat{H}+\epsilon)}
=
\lim_{\epsilon \to 0+}  \left[\frac{1}{-(i\hat{H}+\epsilon)} e^{-T(i\hat{H}+\epsilon)}
\right]_0^{\infty}
=-i\hat{H}^{-1}.
\end{equation}
This fact and \eqref{InverseH} imply
\begin{equation}
\Delta_F(\bar{\bold{E}}, \bold{X}_{\hat{D}_{T}}(\bar{\tau}); \; \bar{\bold{E}},'  \bold{X}_{\hat{D}_{T}}'(\bar{\tau}'))
=
i \int _0^{\infty} dT <\bar{\bold{E}}, \bold{X}_{\hat{D}_{T}}(\bar{\tau})|  e^{-iT\hat{H}} |\bar{\bold{E}},' \bold{X}_{\hat{D}_{T}}'(\bar{\tau}') >.
\end{equation}
In order to define two-point correlation functions that is invariant under the general coordinate transformations in the string geometry, we define in and out states as
\begin{eqnarray}
&&||\bold{X}_i \,|\,\bold{E}_f, ; \bold{E}_i>_{in} \nonumber \\
&:=& \int_{\bold{E}_i}^{\bold{E}_f} \mathcal{D} \bold{E}'|\bar{\bold{E}},' \bold{X}_{\hat{D}_{T}i}:=\bold{X}'(\bar{\tau}'=-\infty)> \nonumber \\
&&<\bold{X}_{\hat{D}_{T}f}\,|\,\bold{E}_f, ; \bold{E}_i||_{out} \nonumber \\
&:=& \int_{\bold{E}_i}^{\bold{E}_f} \mathcal{D} \bold{E} <\bar{\bold{E}}, \bold{X}_{\hat{D}_{T}f}:=\bold{X}_{\hat{D}_{T}}(\bar{\tau}=\infty)|,
\end{eqnarray}
where $\bold{E}_i$ and $\bold{E}_f$ represent the vielbein of the super cylinders at $\bar{\tau}=\pm \infty$, respectively. $\int$ in $\int \mathcal{D} \bold{E}$ includes $\sum_{\mbox{compact topologies}}$, where $\mathcal{D}\bold{E}$ is the invariant measure of the super vielbein $\bold{E}$ on the two-dimensional super Riemannian manifolds $\bold{\Sigma}$.   $\bold{E}$ and $\bar{\bold{E}}$ are related to each others by the super diffeomorphism and the super Weyl transformations. When we insert asymptotic states, we integrate out $\bold{X}_{\hat{D}_{T}f}$, $\bold{X}_{\hat{D}_{T}i}$, $\bold{E}_f$ and $\bold{E}_i$ in the two-point correlation function for these states;
\begin{eqnarray}
\Delta_F(\bold{X}_{\hat{D}_{T}f}; \bold{X}_{\hat{D}_{T}i}|\bold{E}_f, ; \bold{E}_i) := i \int _0^{\infty} dT <\bold{X}_{\hat{D}_{T}f} \,|\,\bold{E}_f, ; \bold{E}_i||_{out}  e^{-iT\hat{H}} ||\bold{X}_{\hat{D}_{T}i} \,|\,\bold{E}_f, ; \bold{E}_i>_{in}.
\end{eqnarray}
By inserting 
\begin{eqnarray}
1&=&
\int d\bar{\bold{E}}_{m}  d\bold{X}_{\hat{D}_{T}Tm} (\bar{\tau}_m ) 
|\bar{\bold{E}}_{m},  \bold{X}_{\hat{D}_{T}Tm}(\bar{\tau}_m) > 
e^{-\tilde{\psi}^{\dagger}_m \cdot \tilde{\psi}_m}
<\bar{\bold{E}}_{m}, \bold{X}_{\hat{D}_{T}Tm}(\bar{\tau}_m) |
\nonumber \\ 
1&=&
\int  dp_{X}^i
| p_{X}^i>
< p_{X}^i|,
\end{eqnarray}
this can be written as in \cite{Sato:2017qhj},
\begin{eqnarray}
&&\Delta_F(\bold{X}_{\hat{D}_{T}f}; \bold{X}_{\hat{D}_{T}i}|\bold{E}_f, ; \bold{E}_i) \nonumber \\
&:=&i\int _0^{\infty} dT <\bold{X}_{\hat{D}_{T}f} \,|\,\bold{E}_f, ; \bold{E}_i||_{out}  e^{-iT\hat{H}} ||\bold{X}_{\hat{D}_{T}i} \,|\,\bold{E}_f, ; \bold{E}_i>_{in}\nonumber \\
&=&i\int _0^{\infty} dT  \lim_{N \to \infty} \int_{\bold{E}_i}^{\bold{E}_f} \mathcal{D} \bold{E} \int_{\bold{E}_i}^{\bold{E}_f} \mathcal{D} \bold{E}' 
\prod_{n=1}^N \int d \bar{\bold{E}}_{n} d\bold{X}_{\hat{D}_{T}n}(\bar{\tau}_n)
e^{-i\psi^\dagger_n\cdot\psi_n}
\nonumber \\
&&\prod_{m=0}^N <\bar{\bold{E}}_{m+1}, \bold{X}_{\hat{D}_{T}m+1}(\bar{\tau}_{m+1})| e^{-i\frac{1}{N}T \hat{H}} |\bar{\bold{E}}_{m}, \bold{X}_{\hat{D}_{T}m}(\bar{\tau}_m),> \nonumber \\
&=&i\int _0^{\infty} dT_0 \lim_{N \to \infty} \int d T_{N+1}
\int_{\bold{E}_i}^{\bold{E}_f}\mathcal{D} \bold{E}  
\prod_{m=1}^N \prod_{i=0}^N
\int d T_m  d\bold{X}_{\hat{D}_{T}m}(\bar{\tau}_m)  e^{-\tilde{\psi}^{\dagger}_m \cdot \tilde{\psi}_m}
\nonumber \\
&&
\int dp_{X}^i
<X_{i+1}| 
p_{X}^i>
<p_{X}^i|
<\tilde{\psi}_{i+1}|
e^{-\frac{1}{N}T_i \hat{H}}
|\tilde{\psi}_{i}> | X_i> \delta(T_{i}-T_{i+1})  \nonumber \\
&=&
i\int _0^{\infty} dT_0 \lim_{N \to \infty} \int d T_{N+1} 
\int_{\bold{E}_i}^{\bold{E}_f}\mathcal{D} \bold{E}
\prod_{m=1}^N \prod_{i=0}^N
\int d T_m  d\bold{X}_{\hat{D}_{T}m}(\bar{\tau}_m)  e^{-\tilde{\psi}^{\dagger}_m \cdot \tilde{\psi}_m}
\nonumber \\
&&
\int  dp_{X}^i
 e^{-\frac{1}{N}T_i H (p_{X}^i, \sqrt{\bar{h}}G_{\mu\nu}(X_i(\bar{\tau}_i))\,\psi_i^\nu(\bar{\tau}_i), 0, \bold{X}_{\hat{D}_{T}i}(\bar{\tau}_{i}), \bar{\bold{E}})}
e^{\tilde{\psi}^{\dagger}_{i+1} \cdot \tilde{\psi}_i}
\delta(T_{i}-T_{i+1}) 
e^{i(p_{X}^i\cdot(X_{i+1}-X_{i}))}
\nonumber \\
&=&i\int _0^{\infty} dT_0 \lim_{N \to \infty} d T_{N+1} \int_{\bold{E}_i}^{\bold{E}_f} \mathcal{D} \bold{E} \prod_{n=1}^N \int d T_n  d\bold{X}_{\hat{D}_{T}n}(\bar{\tau}_n)    \prod_{m=0}^N \int dp_{T_m}  dp_{X_m}(\bar{\tau}_m)  \nonumber \\
&&\exp \Biggl(i\sum_{m=0}^N \Delta t
\Bigl(p_{T_m} \frac{T_{m}-T_{m+1}}{\Delta t} 
+ \int d\bar{\sigma} \bar{e} \, p_{X_m}(\bar{\tau}_m)\frac{X_{m}(\bar{\tau}_m)-X_{m+1}(\bar{\tau}_{m+1})}{\Delta t} 
\nonumber \\
&&+i\,\psi^\dagger_m\cdot\frac{\psi_m(\bar{\tau}_m)
-\psi_{m+1}(\bar{\tau}_{m+1})}{\Delta t}
-T_m H( p_{X_m}(\bar{\tau}_m),  \sqrt{\bar{h}}G_{\mu\nu}(X_{m}(\bar{\tau}_m))\,\psi^\nu_m(\bar{\tau}_m), 0, \bold{X}_{\hat{D}_{T}m}(\bar{\tau}_m), 
\bar{\bold{E}})\Bigr) \Biggr) 
%e^{i\psi^\dagger_{N}\cdot\psi_{N+1}}
\nonumber \\
&=&
i \int_{\bold{E}_i \bold{X}_{\hat{D}_{T} i}}^{\bold{E}_f, \bold{X}_{\hat{D}_{T}f}}  \mathcal{D} \bold{E} \mathcal{D} \bold{X}_{\hat{D}_{T}}(\bar{\tau}) 
\int \mathcal{D} T  
\int 
\mathcal{D} p_T
\mathcal{D}p_{X} (\bar{\tau})
 \nonumber \\
&&
\exp \Biggl(i\int_{-\infty}^{\infty} dt \Bigr(
 p_{T}(t) \frac{d}{dt} T(t) 
+  \int d\bar{\sigma} \bar{e}\, p_{X \mu}(\bar{\tau}(t), t) \frac{d}{dt} X^{\mu}(\bar{\tau}(t), t) \nn\\
&&+ \int d\bar{\sigma} \frac{i}{2} \sqrt{\bar{h}}\, G_{\mu\nu}(X(\bar{\tau}(t), t))\,\bar{\psi}^\mu(\bar{\tau}(t), t)
\,\gamma^0\, \frac{d}{dt}\psi^\nu(\bar{\tau}(t), t)
\nonumber \\
&&
-T(t) H( p_{X}(\bar{\tau}(t), t), \sqrt{\bar{h}}G_{\mu\nu}(X(\bar{\tau}(t), t))\,\psi^\nu(\bar{\tau}(t), t), 0, \bold{X}_{\hat{D}_{T}}(\bar{\tau}(t), t), \bar{\bold{E}})\Bigr) \Biggr),  \label{canonicalform}
\end{eqnarray}
where  $\bar{\bold{E}}_{ 0}=\bar{\bold{E}}'$, $\bold{X}_{\hat{D}_{T}0}(\bar{\tau}_0)=\bold{X}_{\hat{D}_{T}i}$, $\bar{\tau}_0=-\infty$, $\bar{\bold{E}}_{ N+1}=\bar{\bold{E}}$, $\bold{X}_{\hat{D}_{T}N+1}(\bar{\tau}_{N+1})=\bold{X}_{\hat{D}_{T}f}$, $\bar{\tau}_{N+1}=\infty$, and $\Delta t:=\frac{1}{\sqrt{N}}$.
A trajectory of points $[\bar{\bold{\Sigma}}, \bold{X}_{\hat{D}_{T}}(\bar{\tau})]$ is necessarily continuous in $\mathcal{M}_D$ so that the kernel $<\bar{\bold{E}}_{m+1}, \bold{X}_{\hat{D}_{T}m+1}(\bar{\tau}_{m+1})| e^{-i\frac{1}{N}T_m \hat{H}} |\bar{\bold{E}}_{m}, \bold{X}_{\hat{D}_{T}m}(\bar{\tau}_m)>$ in the fourth line is non-zero when $N \to \infty$.

By integrating out $p_{X}(\bar{\tau}(t), t)$, we move from the canonical formalism to the Lagrange formalism.
Because the exponent of (\ref{canonicalform}) is at most the second order in $p_{X}(\bar{\tau}(t), t)$,
integrating out $p_{X}(\bar{\tau}(t), t)$ is equivalent to 
substituting into (\ref{canonicalform}), the solution $p_{X}(\bar{\tau}(t), t)$ of  
\Eqr{
&&-i\bar{e}\,\frac{d}{dt}X^\mu
+iT\bar{e}\,\left[
n^{\bar{\sigma}}\pd_{\bar{\sigma}} X^\mu
+\,i\,\frac{\sqrt{\bar{h}}}
{\bar{e}^2}\,
\pd_{\bar{\sigma}} X^\nu {B_{\nu}}^\mu
\right.
\nn\\
&&\left.~~~
-\frac{i}{2}\,\frac{\bar{h}}
{\bar{e}^2}\,
\bar{\psi}^\nu \left(
{{\Gamma_\nu}^\mu}_\rho(X)+{{H_\nu}^\mu}_\rho(X)\,
\gamma_5\right)\gamma^0\,\psi^\rho
-\frac{1}{2}\,
\,\frac{\bar{h}}
{\bar{e}^2}\,
\left(-2\bar{\chi}_a\,\gamma^0\,\gamma^a\,\psi^\mu
\right)
\right]
\nn\\
&&~~~
+iT\, \sqrt{\bar{h}} p^\mu_X 
-iT\sqrt{\bar{h}}
g^{\mu\nu}(X)
p_{\nu X}=0\,,
  \label{dS'}
}
which is obtained by differentiating  the exponent of (\ref{canonicalform})
with respect to $p_{X}(\bar{\tau}(t), t)$. The solution is given by  
\Eqr{
p_{\mu X}&=&
\frac{1}{T}\frac{\bar{e}}
{\sqrt{\bar{h}}}\,
G_{\mu\nu}\,
\left[\frac{d}{dt}X^\nu
-T\left\{n^{\bar{\sigma}}\pd_{\bar{\sigma}} X^\nu
-\frac{1}{2}
\,\frac{\bar{h}}{\bar{e}^2}\,
\left(-2\bar{\chi}_a\,\gamma^0\,\gamma^a\,\psi^\nu
\right)\right\}\right]\nn\\
&&-\frac{i}{\bar{e}}\left[
\pd_{\bar{\sigma}} X^\gamma B_{\gamma\mu}
-\frac{\sqrt{\bar{h}}}{2}
\bar{\psi}^\kappa \left(
\Gamma_{\kappa\mu\rho}(X)+H_{\kappa\mu\rho}(X)\,
\gamma_5\right)\gamma^0\,\psi^\rho\right],
  \label{pmu}
}
up to the first order in the classical backgrounds $g_{\mu\nu}(X)$ and $B_{\mu \nu}(X)$.
By substituting this, we obtain
\begin{align}
&\Delta_F(\bold{X}_{\hat{D}_{T}f}; \bold{X}_{\hat{D}_{T}i}|\bold{E}_f ; \bold{E}_i) \nonumber \\
&\quad
=i \int_{\bold{E}_i \bold{X}_{\hat{D}_{T}i}}^{\bold{E}_f, \bold{X}_{\hat{D}_{T}f}} 
\mathcal{D} T
\mathcal{D} \bold{E}  \mathcal{D} \bold{X}_{\hat{D}_{T}}(\bar{\tau})
\mathcal{D} p_T 
\nonumber \\
&\qquad\quad
\exp \Biggl(i \int_{-\infty}^{\infty} dt \Bigl( p_{T}(t) \frac{d}{dt} T(t)   \nonumber \\
&\qquad\qquad\qquad
+\int d\bar{\sigma} \sqrt{\bar{h}}G_{\mu\nu}(X(\bar{\tau}(t), t)) ( 
\frac{1}{2}\bar{h}^{00}\frac{1}{T(t)}\partial_{t} X^{\mu}(\bar{\tau}(t), t)\partial_{t} X^{\nu}(\bar{\tau}(t), t) \nonumber \\
&\qquad\qquad\qquad
 +\bar{h}^{01}\partial_{t} X^{\mu}(\bar{\tau}(t), t)\partial_{\bar{\sigma}} X^{\nu}(\bar{\tau}(t), t) 
+\frac{1}{2}\bar{h}^{11}T(t)\partial_{\bar{\sigma}} X^{\mu}(\bar{\tau}(t), t)\partial_{\bar{\sigma}} X^{\nu}(\bar{\tau}(t), t)
)  \nonumber \\
&\qquad\qquad\qquad+\int d\bar{\sigma}\,i\,
B_{\mu\nu} (X(\bar{\tau}(t), t))
\partial_{t} X^{\mu}(\bar{\tau}(t), t)\partial_{\bar{\sigma}} X^{\nu}(\bar{\tau}(t), t) 
\nonumber \\
&\qquad\qquad\qquad
+\frac{1}{2}\int d\bar{\sigma} \,\sqrt{\bar{h}}\,
\left(i\,G_{\mu\nu}\,\bar{\psi}^\mu
\,\gamma^0\,\frac{d}{dt}\psi^\nu
+i\,\bar{\psi}^\nu \left(
\Gamma_{\nu\mu\rho}+H_{\nu\mu\rho}\,
\gamma_5\right)\gamma^0\,\frac{d}{dt}X^\mu(\bar{\tau}(t), t)\,
\psi^\rho
\right.
\nn\\
&\qquad\qquad\qquad\left.
-2\,G_{\mu\nu}\,\bar{\chi}_a\,\gamma^0\,\gamma^a\,\psi^\mu
\frac{d}{dt}X^\nu(\bar{\tau}(t), t)
\right)
\nn\\
&\qquad\qquad\qquad
+\frac{1}{2}\,T
\int d\bar{\sigma}\sqrt{\bar{h}}\,
(iG_{\mu\nu}\,\bar{\psi}^\mu\,\gamma^1\pd_{\bar{\sigma}}\psi^\nu
+i\,\bar{\psi}^\nu \left(
\Gamma_{\nu\mu\rho}+H_{\nu\mu\rho}\,
\gamma_5\right)\gamma^1\,\pd_{\bar{\sigma}} 
X^\mu(\bar{\tau}(t), t)\,
\psi^\rho
\nn\\
&\qquad\qquad\qquad-2\,G_{\mu\nu}\,
\bar{\chi}_a\,\gamma^1\,\gamma^a\,\psi^\mu
\pd_{\bar{\sigma}} X^\nu(\bar{\tau}(t), t)
+\frac{1}{2}G_{\mu\nu}\,\bar{\psi}^\mu\,\psi^\nu\,
\bar{\chi}_a\,\gamma^b\,\gamma^a \chi_b
\nn\\
&\qquad\qquad\qquad
+\frac{1}{6}R_{\mu\nu\lambda\rho}\,
\bar{\psi}^\mu\,\psi^\lambda\,
\bar{\psi}^\nu\,\psi^\rho
-\frac{i}{3}H_{\mu\nu\rho}\,\bar{\chi}_a\,
\gamma^b\,\gamma^a\,\bar{\psi}^\mu\,\psi^\nu\,
\gamma_b\gamma_5\psi^\rho
\nn\\
&\qquad\qquad\qquad
-\frac{1}{4}D_\rho H_{\mu\nu\lambda}
\bar{\psi}^\mu\,\psi^\rho\,\bar{\psi}^\lambda\,
\gamma_5\psi^\nu\,+R_{\bar{h}} \frac{\lambda}{2\pi}
%+\frac{1}{4}G^{\alpha\beta}\,H_{\mu\nu\alpha}\,
%H_{\lambda\rho\beta}\,\bar{\psi}^\mu\,\gamma_5\,\psi^\nu\,
%\bar{\psi}^\lambda\,\gamma_5\psi^\rho
)
\Bigr) \Biggr), \label{pathint1}
\end{align}
where we use  
the ADM decomposition of the two-dimensional metric,
\Eqr{
\bar{h}_{mn}=
\left(
\begin{array}{cc}
\bar{n}^2+ \bar{n}_{\bar{\sigma}} \bar{n}^{\bar{\sigma}} & \bar{n}_{\bar{\sigma}} \\
\bar{n}_{\bar{\sigma}} & \bar{e}^2
\end{array}
\right)\,,
~~~~~~
\sqrt{\bar{h}}=\bar{n}\bar{e}\,,
~~~~~~\bar{h}^{mn}=
\begin{pmatrix}
\frac{1}{\bar{n}^2} & -\frac{\bar{n}^{\bar{\sigma}}}
{\bar{n}^2} \\
-\frac{\bar{n}^{\bar{\sigma}}}{\bar{n}^2} &
\bar{e}^{-2}+\left(
\frac{\bar{n}^{\bar{\sigma}}}{\bar{n}}\right)^2 \label{ADM}
\end{pmatrix}\,,
}
$\gamma^1=\left(
\begin{array}{cc}
0 & 1 \\
1 & 0
\end{array}
\right)$, and (\ref{NoOnshell}) by choosing a function $\omega ({\bf X}_{\hat{D}_{T}})$ as
\Eqr{
&&\omega ({\bf X}_{\hat{D}_{T}})=
\int d\bar{\sigma}
\sqrt{\bar{h}}\, \Biggl(R_{\bar{h}} \frac{\lambda}{4\pi}
+\frac{1}{\bar{e}^2}G_{\mu\nu}
\pd_{\bar{\sigma}} {X}^\mu \pd_{\bar{\sigma}} {X}^\nu
+iG_{\nu\mu}\,\bar{\psi}^\mu\,\gamma^1\pd_{\bar{\sigma}}\psi^\nu
\nn\\
&&~
-\frac{i}{2}\sqrt{\bar{h}}
\bar{\psi}^\nu\left(
\Gamma_{\nu\mu\rho}+H_{\nu\mu\rho}\,
\gamma_5\right)\gamma^0\,\psi^\rho
\left\{-2\,n^{\bar{\sigma}}\,\pd_{\bar{\sigma}} X^\mu
-\frac{\bar{h}}{\bar{e}^2}
\left(-2\bar{\chi}_a\,\gamma^0\,\gamma^a\,\psi^\mu\right)
\right\}
\nn\\
&&~~~
+i\,\bar{\psi}^\nu \left(
\Gamma_{\nu\mu\rho}+H_{\nu\mu\rho}\,
\gamma_5\right)\gamma^1\,\pd_{\bar{\sigma}} X^\mu\,
\psi^\rho
-2\,G_{\mu\nu}\,\bar{\chi}_a\,\gamma^1\,\gamma^a\,\psi^\mu
\pd_{\bar{\sigma}} X^\nu
\nn\\
&&~
-i\,\sqrt{\bar{h}}
\left(-2\bar{\chi}_a\,\gamma^0\,\gamma^a\,\psi^\mu
\right)\left(iG_{\mu\nu}n^{\bar{\sigma}}\,\pd_{\bar{\sigma}} X^\nu
+\frac{\sqrt{\bar{h}}}{\bar{e}^2}\pd_{\bar{\sigma}} X^\lambda
B_{\lambda\mu}\right)
\nn\\
&&~-\frac{i}{4}\int d\bar{\sigma}\,\frac{\bar{h}}
{\bar{e}^2}\,i\,\bar{\psi}^\mu \partial_{\nu}{{H_\mu}^\nu}_\rho(X)\,
\gamma_5\,\gamma^0\,\psi^\rho \nn \\
&&~-\frac{1}{4}\,\frac{\bar{h}}{e^2}
G_{\mu\nu}\left(-2\bar{\chi}_a\,\gamma^0\,\gamma^a\,\psi^\mu
\right)\left(-2\bar{\chi}_b\,\gamma^0\,\gamma^b\,\psi^\nu
\right)
\nn\\
&&~
+\frac{1}{2}G_{\mu\nu}\,\bar{\psi}^\mu\,\psi^\nu\,
\bar{\chi}_a\,\gamma^b\,\gamma^a \chi_b
+\frac{1}{6}R_{\mu\nu\lambda\rho}\,
\bar{\psi}^\mu\,\psi^\lambda\,
\bar{\psi}^\nu\,\psi^\rho
-\frac{i}{3}H_{\mu\nu\rho}\,\bar{\chi}_a\,
\gamma^b\,\gamma^a\,\bar{\psi}^\mu\,\psi^\nu\,
\gamma_b\gamma_5\psi^\rho
\nn\\
&&~
-\frac{1}{4}D_\rho H_{\mu\nu\lambda}
\bar{\psi}^\mu\,\psi^\rho\,\bar{\psi}^\lambda\,
\gamma_5\psi^\nu\,
%+\frac{1}{4}G^{\alpha\beta}\,H_{\mu\nu\alpha}\,
%H_{\lambda\rho\beta}(X)\,\bar{\psi}^\mu\,\gamma_5\,\psi^\nu\,
%\bar{\psi}^\lambda\,\gamma_5\psi^\rho\,
\Biggr),
  \label{f'}
}
where $R_{\bar{h}}$ is the scalar curvature of the two-dimensional metric $\bar{h}_{mn}$ and $\lambda$ is a constant, which will be identified with the logarithm of the string coupling constant $g_s$. In the following, we consider only constant dilaton backgrounds because it is not known a world-sheet theory of  superstrings in non-constant dilaton backgrounds.

In this way, the Green function can generate all the terms without $\bar{\tau}$ derivatives in the string action as in (\ref{NoOnshell}), but cannot do those with $\bar{\tau}$ derivatives,   which need  to be derived non-trivially, because the coordinates  $ X^{\mu}(\bar{\tau})$ in string geometry theory are defined on the $\bar{\tau}$ constant lines.  
We should note that the time derivative in \eqref{pathint1} is in terms of $t$, not $\bar{\tau}$ at this moment. In the following, we will see that $t$ can be fixed to $\bar{\tau}$ by using a reparametrization of $t$ that parametrizes a trajectory.

By inserting
$\int \mathcal{D}c \mathcal{D}b
e^{\int_0^{1} dt \left(\frac{d b(t)}{dt} \frac{d c(t)}{dt}\right)
},$
where $b(t)$ and $c(t)$ are $bc$-ghost, we obtain 
\begin{align}
&\Delta_F(\bold{X}_{\hat{D}_{T}f}; \bold{X}_{\hat{D}_{T}i}|\bold{E}_f ; \bold{E}_i) \nonumber \\
&\quad
=Z_0 \int_{\bold{E}_i \bold{X}_{\hat{D}_{T}i}}^{\bold{E}_f, \bold{X}_{\hat{D}_{T}f}} 
\mathcal{D} T
\mathcal{D} \bold{E}  \mathcal{D} \bold{X}_{\hat{D}_{T}}(\bar{\tau})
\mathcal{D} p_T
\mathcal{D}c \mathcal{D}b  \nonumber \\
&\qquad\quad
\exp \Biggl(- \int_{-\infty}^{\infty} dt \Bigl(
-i p_{T}(t) \frac{d}{dt} T(t)  
 +\frac{d b(t)}{dt} \frac{d (T(t) c(t))}{dt}\nonumber \\
&\qquad\qquad\qquad
+\int d\bar{\sigma} \sqrt{\bar{h}}G_{\mu\nu}(X(\bar{\tau}(t), t))  ( 
\frac{1}{2}\bar{h}^{00}\frac{1}{T(t)}\partial_{t} X^{\mu}(\bar{\tau}(t), t)\partial_{t} X^{\nu}(\bar{\tau}(t), t) \nonumber \\
&\qquad\qquad\qquad
+\bar{h}^{01}\partial_{t} X^{\mu}(\bar{\tau}(t), t)\partial_{\bar{\sigma}} X^{\nu}(\bar{\tau}(t), t) 
+\frac{1}{2}\bar{h}^{11}T(t)\partial_{\bar{\sigma}} X^{\mu}(\bar{\tau}(t), t)\partial_{\bar{\sigma}} X^{\nu}(\bar{\tau}(t), t)
)  \nonumber \\
&\qquad\qquad\qquad+\int d\bar{\sigma}\,i\,
B_{\mu\nu} (X(\bar{\tau}(t), t))
\partial_{t} X^{\mu}(\bar{\tau}(t), t)\partial_{\bar{\sigma}} X^{\nu}(\bar{\tau}(t), t) 
\nn\\
&\qquad\qquad\qquad
+\frac{1}{2}\int d\bar{\sigma} \,\sqrt{\bar{h}}\,
\left(i\,G_{\mu\nu}\,\bar{\psi}^\mu
\,\gamma^0\,\frac{d}{dt}\psi^\nu
+i\,\bar{\psi}^\nu \left(
\Gamma_{\nu\mu\rho}+H_{\nu\mu\rho}\,
\gamma_5\right)\gamma^0\,\frac{d}{dt}X^\mu(\bar{\tau}(t), t)\,
\psi^\rho
\right.
\nn\\
&\qquad\qquad\qquad\left.
-2\,G_{\mu\nu}\,\bar{\chi}_a\,\gamma^0\,\gamma^a\,\psi^\mu
\frac{d}{dt}X^\nu(\bar{\tau}(t), t)
\right)
\nn\\
&\qquad\qquad\qquad
+\frac{1}{2}\,T
\int d\bar{\sigma}\sqrt{\bar{h}}\,
\left(
iG_{\mu\nu}\,\bar{\psi}^\mu\,\gamma^1\pd_{\bar{\sigma}}\psi^\nu
+i\,\bar{\psi}^\nu \left(
\Gamma_{\nu\mu\rho}+H_{\nu\mu\rho}\,
\gamma_5\right)\gamma^1\,\pd_{\bar{\sigma}} X^\mu(\bar{\tau}(t), t)\,
\psi^\rho\right.
\nn\\
&\qquad\qquad\qquad-2\,G_{\mu\nu}\,
\bar{\chi}_a\,\gamma^1\,\gamma^a\,\psi^\mu
\pd_{\bar{\sigma}} X^\nu(\bar{\tau}(t), t)
+\frac{1}{2}G_{\mu\nu}\,\bar{\psi}^\mu\,\psi^\nu\,
\bar{\chi}_a\,\gamma^b\,\gamma^a \chi_b
\nn\\
&\qquad\qquad\qquad
+\frac{1}{6}R_{\mu\nu\lambda\rho}\,
\bar{\psi}^\mu\,\psi^\lambda\,
\bar{\psi}^\nu\,\psi^\rho
-\frac{i}{3}H_{\mu\nu\rho}\,\bar{\chi}_a\,
\gamma^b\,\gamma^a\,\bar{\psi}^\mu\,\psi^\nu\,
\gamma_b\gamma_5\psi^\rho
\nn\\
&\qquad\qquad\qquad\left.
-\frac{1}{4}D_\rho H_{\mu\nu\lambda}
\bar{\psi}^\mu\,\psi^\rho\,\bar{\psi}^\lambda\,
\gamma_5\psi^\nu\,+R_{\bar{h}} \frac{\lambda}{2\pi}
%+\frac{1}{4}G^{\alpha\beta}\,H_{\mu\nu\alpha}\,
%H_{\lambda\rho\beta}\,\bar{\psi}^\mu\,\gamma_5\,\psi^\nu\,
%\bar{\psi}^\lambda\,\gamma_5\psi^\rho
\right)
\Bigr) \Biggr), 
\label{PropWMult}
\end{align}
where we redefine as $c(t) \to T(t) c(t)$, and $Z_0$ represents an overall constant factor. In the following, we will rename it $Z_1, Z_2, \cdots$ when the factor changes. The integrand variable $p_T (t)$ plays the role of the Lagrange multiplier providing the following condition,
\begin{align}
F_1(t):=\frac{d}{dt}T(t)=0,
\label{F1gauge}
\end{align}
which can be understood as a gauge fixing condition. Indeed, by choosing this gauge in
\begin{align}
&\Delta_F(\bold{X}_{\hat{D}_{T}f}; \bold{X}_{\hat{D}_{T}i}|\bold{E}_f ; \bold{E}_i) \nonumber \\
&\quad
=Z_1 \int_{\bold{E}_i \bold{X}_{\hat{D}_{T}i}}^{\bold{E}_f, \bold{X}_{\hat{D}_{T}f}} 
\mathcal{D} T
\mathcal{D} \bold{E}  \mathcal{D} \bold{X}_{\hat{D}_{T}}(\bar{\tau})
\nonumber \\
&\qquad\quad
\exp \Biggl(- \int_{-\infty}^{\infty} dt \Bigl(
\int d\bar{\sigma} \sqrt{\bar{h}} G_{\mu\nu}(X(\bar{\tau}(t), t)) ( 
\frac{1}{2}\bar{h}^{00}\frac{1}{T(t)}\partial_{t} X^{\mu}(\bar{\tau}(t), t)\partial_{t} X^{\nu}(\bar{\tau}(t), t)
\nonumber \\
&\qquad\qquad\qquad
 +\bar{h}^{01}\partial_{t} X^{\mu}(\bar{\tau}(t), t)\partial_{\bar{\sigma}} X^{\nu}(\bar{\tau}(t), t) 
+\frac{1}{2}\bar{h}^{11}T(t)\partial_{\bar{\sigma}} X^{\mu}(\bar{\tau}(t), t)\partial_{\bar{\sigma}} X^{\nu}(\bar{\tau}(t), t)
)  \nonumber \\
&\qquad\qquad\qquad+\int d\bar{\sigma}\,i\,
B_{\mu\nu} (X(\bar{\tau}(t), t))
\partial_{t} X^{\mu}(\bar{\tau}(t), t)\partial_{\bar{\sigma}} X^{\nu}(\bar{\tau}(t), t) 
\nonumber \\
&\qquad\qquad\qquad
+\frac{1}{2}\int d\bar{\sigma} \,\sqrt{\bar{h}}\,
\left(i\,G_{\mu\nu}\,\bar{\psi}^\mu
\,\gamma^0\,\frac{d}{dt}\psi^\nu
+i\,\bar{\psi}^\nu \left(
\Gamma_{\nu\mu\rho}+H_{\nu\mu\rho}\,
\gamma_5\right)\gamma^0\,\frac{d}{dt}X^\mu(\bar{\tau}(t), t)\,
\psi^\rho
\right.
\nn\\
&\qquad\qquad\qquad\left.
-2\,G_{\mu\nu}\,\bar{\chi}_a\,\gamma^0\,\gamma^a\,\psi^\mu
\frac{d}{dt}X^\nu(\bar{\tau}(t), t)
\right)
\nn\\
&\qquad\qquad\qquad
+\frac{1}{2}\,T(t)
\int d\bar{\sigma}\sqrt{\bar{h}}\,
(iG_{\mu\nu}\,\bar{\psi}^\mu\,\gamma^1\pd_{\bar{\sigma}}\psi^\nu
+i\,\bar{\psi}^\nu \left(
\Gamma_{\nu\mu\rho}+H_{\nu\mu\rho}\,
\gamma_5\right)\gamma^1\,\pd_{\bar{\sigma}} X^\mu(\bar{\tau}(t), t)\,
\psi^\rho
\nn\\
&\qquad\qquad\qquad-2\,G_{\mu\nu}\,
\bar{\chi}_a\,\gamma^1\,\gamma^a\,\psi^\mu
\pd_{\bar{\sigma}} X^\nu(\bar{\tau}(t), t)
+\frac{1}{2}G_{\mu\nu}\,\bar{\psi}^\mu\,\psi^\nu\,
\bar{\chi}_a\,\gamma^b\,\gamma^a \chi_b
\nn\\
&\qquad\qquad\qquad
+\frac{1}{6}R_{\mu\nu\lambda\rho}\,
\bar{\psi}^\mu\,\psi^\lambda\,
\bar{\psi}^\nu\,\psi^\rho
-\frac{i}{3}H_{\mu\nu\rho}\,\bar{\chi}_a\,
\gamma^b\,\gamma^a\,\bar{\psi}^\mu\,\psi^\nu\,
\gamma_b\gamma_5\psi^\rho
\nn\\
&\qquad\qquad\qquad
-\frac{1}{4}D_\rho H_{\mu\nu\lambda}
\bar{\psi}^\mu\,\psi^\rho\,\bar{\psi}^\lambda\,
\gamma_5\psi^\nu\, +R_{\bar{h}} \frac{\lambda}{2\pi}
%+\frac{1}{4}G^{\alpha\beta}\,H_{\mu\nu\alpha}\,
%H_{\lambda\rho\beta}\,\bar{\psi}^\mu\,\gamma_5\,\psi^\nu\,
%\bar{\psi}^\lambda\,\gamma_5\psi^\rho
)
\Bigr) \Biggr),
\label{pathint2}
\end{align}
we obtain \eqref{PropWMult}.
The expression \eqref{pathint2} has a manifest one-dimensional diffeomorphism symmetry with respect to $t$, where $T(t)$ is transformed as an einbein \cite{Schwinger0}. 

Under $\frac{d\bar{\tau}}{d\bar{\tau}'}=T(t)$, which implies
\begin{eqnarray}
\bar{h}^{00}&=&T^2\bar{h}^{'00} \nonumber \\
\bar{h}^{01}&=&T\bar{h}^{'01} \nonumber \\
\bar{h}^{11}&=&\bar{h}^{'11} \nonumber \\
\sqrt{\bar{h}}&=&\frac{1}{T}\sqrt{\bar{h}'} 
\end{eqnarray}

$T(t)$ disappears in \eqref{pathint2} and we obtain 
\begin{align}
&\Delta_F(\bold{X}_{\hat{D}_{T}f}; \bold{X}_{\hat{D}_{T}i}|\bold{E}_f ; \bold{E}_i) \nonumber \\
&=
Z_2 \int_{\bold{E}_i \bold{X}_{\hat{D}_{T}i}}^{\bold{E}_f, \bold{X}_{\hat{D}_{T}f}} 
\mathcal{D} \bold{E}  \mathcal{D} \bold{X}_{\hat{D}_{T}}(\bar{\tau})
\nonumber \\
&\qquad
\exp \Biggl(- \int_{-\infty}^{\infty} dt \Bigl(
\int d\bar{\sigma} \sqrt{\bar{h}}G_{\mu\nu}(X(\bar{\tau}(t), t))  ( 
\frac{1}{2}\bar{h}^{00}\partial_{t} X^{\mu}(\bar{\tau}(t), t)\partial_{t} X^{\nu}(\bar{\tau}(t), t) 
\nonumber \\ &+\bar{h}^{01}\partial_{t} X^{\mu}(\bar{\tau}(t), t)\partial_{\bar{\sigma}} X^{\nu}(\bar{\tau}(t), t) 
+\frac{1}{2}\bar{h}^{11}\partial_{\bar{\sigma}} X^{\mu}(\bar{\tau}(t), t)\partial_{\bar{\sigma}} X^{\nu}(\bar{\tau}(t), t)
)  \nonumber \\
&+\int d\bar{\sigma}\,i\,
B_{\mu\nu} (X(\bar{\tau}(t), t))
\partial_{t} X^{\mu}(\bar{\tau}(t), t)\partial_{\bar{\sigma}} X^{\nu}(\bar{\tau}(t), t) 
\nonumber \\
&
+\frac{1}{2}\int d\bar{\sigma} \,\sqrt{\bar{h}}\,
\left(i\,G_{\mu\nu}\,\bar{\psi}^\mu
\,\gamma^0\,\pd_{\bar{\tau}}\psi^\nu
+i\,\bar{\psi}^\nu \left(
\Gamma_{\nu\mu\rho}+H_{\nu\mu\rho}\,
\gamma_5\right)\gamma^0\,\pd_{\bar{\tau}}X^\mu(\bar{\tau}(t), t)\,
\psi^\rho
\right.
\nn\\
&\qquad\qquad\qquad\left.
-2\,G_{\mu\nu}\,\bar{\chi}_a\,\gamma^0\,\gamma^a\,\psi^\mu
\pd_{\bar{\tau}}X^\nu(\bar{\tau}(t), t)
\right)
\nn\\
&
+\frac{1}{2}\,
\int d\bar{\sigma}\sqrt{\bar{h}}\,
(iG_{\mu\nu}\,\bar{\psi}^\mu\,\gamma^1\pd_{\bar{\sigma}}\psi^\nu
+i\,\bar{\psi}^\nu \left(
\Gamma_{\nu\mu\rho}+H_{\nu\mu\rho}\,
\gamma_5\right)\gamma^1\,\pd_{\bar{\sigma}} 
X^\mu(\bar{\tau}(t), t)\,
\psi^\rho
\nn\\
&\qquad\qquad\qquad-2\,G_{\mu\nu}\,
\bar{\chi}_a\,\gamma^1\,\gamma^a\,\psi^\mu
\pd_{\bar{\sigma}} X^\nu(\bar{\tau}(t), t)
+\frac{1}{2}G_{\mu\nu}\,\bar{\psi}^\mu\,\psi^\nu\,
\bar{\chi}_a\,\gamma^b\,\gamma^a \chi_b
\nn\\
&\qquad\qquad\qquad
+\frac{1}{6}R_{\mu\nu\lambda\rho}\,
\bar{\psi}^\mu\,\psi^\lambda\,
\bar{\psi}^\nu\,\psi^\rho
-\frac{i}{3}H_{\mu\nu\rho}\,\bar{\chi}_a\,
\gamma^b\,\gamma^a\,\bar{\psi}^\mu\,\psi^\nu\,
\gamma_b\gamma_5\psi^\rho
\nn\\
&\qquad\qquad\qquad
-\frac{1}{4}D_\rho H_{\mu\nu\lambda}
\bar{\psi}^\mu\,\psi^\rho\,\bar{\psi}^\lambda\,
\gamma_5\psi^\nu\, +R_{\bar{h}} \frac{\lambda}{2\pi}
%+\frac{1}{4}G^{\alpha\beta}\,H_{\mu\nu\alpha}\,
%H_{\lambda\rho\beta}\,\bar{\psi}^\mu\,\gamma_5\,\psi^\nu\,
%\bar{\psi}^\lambda\,\gamma_5\psi^\rho
)
\Bigr) \Biggr)\,. \label{pathint3}
\end{align}
This action is still invariant under the diffeomorphism with respect to t if $\bar{\tau}$ transforms in the same way as $t$. 

\begin{figure}[htb]
\centering
\includegraphics[width=6cm]{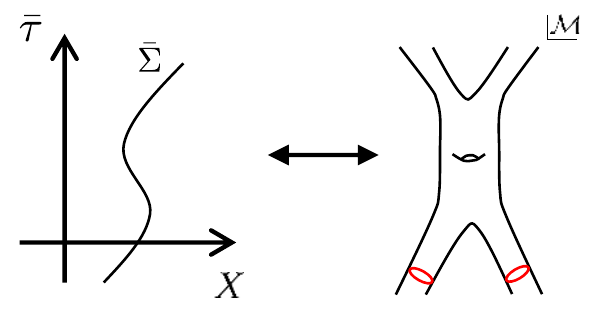}
\caption{A path and a  super Riemann surface. The line on the left is a trajectory in the path integral. The trajectory parametrized by $\bar{\tau}$ from $-\infty$ to $\infty$, represents a super Riemann surface with fixed punctures in $\mathcal{M}$ on the right.} 
\label{Pathintegral}
\end{figure}

If we choose a different gauge
\begin{equation}
F_2(t):=\bar{\tau}(t)-t=0, \label{F2gauge}
\end{equation} 
in \eqref{pathint3}, we obtain 
\begin{align}
&\Delta_F(\bold{X}_{\hat{D}_{T}f}; \bold{X}_{\hat{D}_{T}i}|\bold{E}_f ; \bold{E}_i) \nonumber \\
&\quad
=Z_3 \int_{\bold{E}_i \bold{X}_{\hat{D}_{T}i}}^{\bold{E}_f, \bold{X}_{\hat{D}_{T}f}} 
\mathcal{D} \bold{E}  \mathcal{D} \bold{X}_{\hat{D}_{T}}(\bar{\tau})
\mathcal{D} \alpha \mathcal{D}c \mathcal{D}b
\nonumber \\
&\qquad\quad
\exp \Biggl(- \int_{-\infty}^{\infty} dt \Bigl(\alpha(t) (\bar{\tau}-t) +b(t)c(t)(1-\frac{d \bar{\tau}(t)}{dt})   \nonumber \\
&\qquad\qquad\qquad
+\int d\bar{\sigma} \sqrt{\bar{h}}G_{\mu\nu}(X(\bar{\tau}(t), t))  ( 
\frac{1}{2}\bar{h}^{00}\partial_{t} X^{\mu}(\bar{\tau}(t), t)\partial_{t} X^{\nu}(\bar{\tau}(t), t) 
\nonumber \\
&\qquad\qquad\qquad
+\bar{h}^{01}\partial_{t} X^{\mu}(\bar{\tau}(t), t)\partial_{\bar{\sigma}} X^{\nu}(\bar{\tau}(t), t) 
+\frac{1}{2}\bar{h}^{11}\partial_{\bar{\sigma}} X^{\mu}(\bar{\tau}(t), t)\partial_{\bar{\sigma}} X^{\nu}(\bar{\tau}(t), t)
)  \nonumber \\
&\qquad\qquad\qquad+\int d\bar{\sigma}\,i\,
B_{\mu\nu} (X(\bar{\tau}(t), t))
\partial_{t} X^{\mu}(\bar{\tau}(t), t)\partial_{\bar{\sigma}} X^{\nu}(\bar{\tau}(t), t) 
\nonumber \\
&\qquad\qquad\qquad
+\frac{1}{2}\int d\bar{\sigma} \,\sqrt{\bar{h}}\,
\Bigl(i\,G_{\mu\nu}\,\bar{\psi}^\mu
\,\gamma^a\,\pd_a\psi^\nu
+i\,\bar{\psi}^\nu \left(
\Gamma_{\nu\mu\rho}+H_{\nu\mu\rho}\,
\gamma_5\right)\gamma^a\,\pd_a X^\mu(\bar{\tau}(t), t)\,
\psi^\rho
\nn\\
&\qquad\qquad\qquad
-2\,G_{\mu\nu}\,\bar{\chi}_a\,\gamma^b\,\gamma^a\,\psi^\mu
\pd_bX^\nu(\bar{\tau}(t), t)
+\frac{1}{2}G_{\mu\nu}\,\bar{\psi}^\mu\,\psi^\nu\,
\bar{\chi}_a\,\gamma^b\,\gamma^a \chi_b
\nn\\
&\qquad\qquad\qquad
+\frac{1}{6}R_{\mu\nu\lambda\rho}\,
\bar{\psi}^\mu\,\psi^\lambda\,
\bar{\psi}^\nu\,\psi^\rho
-\frac{i}{3}H_{\mu\nu\rho}\,\bar{\chi}_a\,
\gamma^b\,\gamma^a\,\bar{\psi}^\mu\,\psi^\nu\,
\gamma_b\gamma_5\psi^\rho
\nn\\
&\qquad\qquad\qquad
-\frac{1}{4}D_\rho H_{\mu\nu\lambda}
\bar{\psi}^\mu\,\psi^\rho\,\bar{\psi}^\lambda\,
\gamma_5\psi^\nu\, +R_{\bar{h}} \frac{\lambda}{2\pi}
%+\frac{1}{4}G^{\alpha\beta}\,H_{\mu\nu\alpha}\,
%H_{\lambda\rho\beta}\,\bar{\psi}^\mu\,\gamma_5\,\psi^\nu\,
%\bar{\psi}^\lambda\,\gamma_5\psi^\rho
\Bigr)
\Bigr) \Biggr) \nonumber \\
&\quad
=Z\int_{\bold{E}_i, \bold{X}_{\hat{D}_{T}i}}^{\bold{E}_f, \bold{X}_{\hat{D}_{T}f}} 
\mathcal{D} \bold{E}  \mathcal{D} \bold{X}_{\hat{D}_{T}}
\nonumber \\
&\qquad\quad
\exp \Biggl(- \int_{-\infty}^{\infty} d\bar{\tau} 
\int d\bar{\sigma} \sqrt{\bar{h}}G_{\mu\nu}(X(\bar{\tau}(t), t))  ( 
\frac{1}{2}\bar{h}^{00}\partial_{\bar{\tau}} X^{\mu}(\bar{\sigma}, \bar{\tau})\partial_{\bar{\tau}} X^{\nu}(\bar{\sigma}, \bar{\tau})
 \nonumber \\
&\qquad\qquad\qquad\qquad
+\bar{h}^{01}\partial_{\bar{\tau}} X^{\mu}(\bar{\sigma}, \bar{\tau})\partial_{\bar{\sigma}} X^{\nu}(\bar{\sigma}, \bar{\tau}) 
+\frac{1}{2}\bar{h}^{11}\partial_{\bar{\sigma}} X^{\mu}(\bar{\sigma}, \bar{\tau})\partial_{\bar{\sigma}} X^{\nu}(\bar{\sigma}, \bar{\tau}))
 \nonumber \\
&\qquad\qquad\qquad+\int d\bar{\sigma}\,i\,
B_{\mu\nu} (X(\bar{\sigma}, \bar{\tau}))
\partial_{\bar{\tau}} X^{\mu}(\bar{\sigma}, \bar{\tau})\partial_{\bar{\sigma}} X^{\nu}(\bar{\sigma}, \bar{\tau}) 
\nonumber \\
&\qquad\qquad\qquad
+\frac{1}{2}\int d\bar{\sigma} \,\sqrt{\bar{h}}\,
\Bigl(i\,G_{\mu\nu}\,\bar{\psi}^\mu
\,\gamma^a\,\pd_a\psi^\nu
+i\,\bar{\psi}^\nu \left(
\Gamma_{\nu\mu\rho}+H_{\nu\mu\rho}\,
\gamma_5\right)\gamma^a\,\pd_a X^\mu(\bar{\tau}(t), t)\,
\psi^\rho
\nn\\
&\qquad\qquad\qquad
-2\,G_{\mu\nu}\,\bar{\chi}_a\,\gamma^b\,\gamma^a\,\psi^\mu
\pd_bX^\nu(\bar{\tau}(t), t)
+\frac{1}{2}G_{\mu\nu}\,\bar{\psi}^\mu\,\psi^\nu\,
\bar{\chi}_a\,\gamma^b\,\gamma^a \chi_b
\nn\\
&\qquad\qquad\qquad
+\frac{1}{6}R_{\mu\nu\lambda\rho}\,
\bar{\psi}^\mu\,\psi^\lambda\,
\bar{\psi}^\nu\,\psi^\rho
-\frac{i}{3}H_{\mu\nu\rho}\,\bar{\chi}_a\,
\gamma^b\,\gamma^a\,\bar{\psi}^\mu\,\psi^\nu\,
\gamma_b\gamma_5\psi^\rho
\nn\\
&\qquad\qquad\qquad
-\frac{1}{4}D_\rho H_{\mu\nu\lambda}
\bar{\psi}^\mu\,\psi^\rho\,\bar{\psi}^\lambda\,
\gamma_5\psi^\nu\, +R_{\bar{h}} \frac{\lambda}{2\pi}
%+\frac{1}{4}G^{\alpha\beta}\,H_{\mu\nu\alpha}\,
%H_{\lambda\rho\beta}\,\bar{\psi}^\mu\,\gamma_5\,\psi^\nu\,
%\bar{\psi}^\lambda\,\gamma_5\psi^\rho
\Bigr)
\Biggr). \label{prelastaction}
\end{align}
The path integral is defined over all possible two-dimensional super Riemannian manifolds with fixed punctures in the manifold $\mathcal{M}$ defined by the metric $G_{\mu\nu}$, as in Fig. \ref{Pathintegral}.
The super diffeomorphism times super Weyl invariance of the action in \eqref{prelastaction} implies that the correlation function is given by 
\begin{equation}
\Delta_F(\bold{X}_{\hat{D}_{T}f}; \bold{X}_{\hat{D}_{T}i}|\bold{E}_f ; \bold{E}_i)
=
Z
\int_{\bold{E}_i, \bold{X}_{\hat{D}_{T}i}}^{\bold{E}_f, \bold{X}_{\hat{D}_{T}f}} 
\mathcal{D} \bold{E}  \mathcal{D} \bold{X}_{\hat{D}_{T}}
e^{-\lambda \chi}e^{-S_{s}}, 
\label{FinalPropagator}
\end{equation}
where
\begin{eqnarray}
S_{s}
&=&
\frac{1}{2}\int_{-\infty}^{\infty} d\tau \int d\sigma \sqrt{h(\sigma, \tau)}\biggl(\bigl( h^{mn} (\sigma, \tau) G_{\mu\nu}(X(\sigma, \tau))
+i \varepsilon^{mn}(\sigma, \tau)
B_{\mu\nu}(X(\sigma, \tau))\bigr)
\nn\\
&&~~~\times
 \partial_m X^{\mu}(\sigma, \tau) \partial_n X^{\nu}(\sigma, \tau) 
\nonumber \\
&&~~~
+i\,G_{\mu\nu}\,\bar{\psi}^\mu
\,\gamma^a\,\pd_a\psi^\nu
+i\,\bar{\psi}^\nu \left(
\Gamma_{\nu\mu\rho}+H_{\nu\mu\rho}\,
\gamma_5\right)\gamma^a\,\pd_a X^\mu(\sigma, \tau)\,
\psi^\rho
\nn\\
&&~~~
-2\,G_{\mu\nu}\,\bar{\chi}_a\,\gamma^b\,\gamma^a\,\psi^\mu
\pd_bX^\nu(\sigma, \tau)
+\frac{1}{2}G_{\mu\nu}\,\bar{\psi}^\mu\,\psi^\nu\,
\bar{\chi}_a\,\gamma^b\,\gamma^a \chi_b
\nn\\
&&~~~
+\frac{1}{6}R_{\mu\nu\lambda\rho}\,
\bar{\psi}^\mu\,\psi^\lambda\,
\bar{\psi}^\nu\,\psi^\rho
-\frac{i}{3}H_{\mu\nu\rho}\,\bar{\chi}_a\,
\gamma^b\,\gamma^a\,\bar{\psi}^\mu\,\psi^\nu\,
\gamma_b\gamma_5\psi^\rho
\nn\\
&&~~~
-\frac{1}{4}D_\rho H_{\mu\nu\lambda}
\bar{\psi}^\mu\,\psi^\rho\,\bar{\psi}^\lambda\,
\gamma_5\psi^\nu\,
%+\frac{1}{4}G^{\alpha\beta}\,H_{\mu\nu\alpha}\,
%H_{\lambda\rho\beta}\,\bar{\psi}^\mu\,\gamma_5\,\psi^\nu\,
%\bar{\psi}^\lambda\,\gamma_5\psi^\rho
\biggr),  \label{last}
\end{eqnarray}
and $\chi$ is the Euler number of the two-dimensional Riemannian manifold.
For regularization, we divide the correlation function by $Z$ and the volume of the super diffeomorphism and the super Weyl transformation $V_{diff \times Weyl} $, by renormalizing $\tilde{\phi}$. \eqref{FinalPropagator} is the path-integrals of   perturbative superstrings on an arbitrary background that possess the supermoduli  in the type IIA, type IIB and SO(32) type I  superstring theory for $T=$ IIA, IIB and I, respectively\cite{Bergshoeff:1985qr,  Uematsu}. Especially, in string geometry, the consistency of the perturbation theory around the background \eqref{Gh},  (\ref{Bdmu}),  (\ref{Bmunu}), (\ref{Phi}) and  (\ref{A=0}) determines $d=10$ (the critical dimension).

\section{Conclusion and Discussion}
\setcounter{equation}{0}
In this paper, in the string geometry theory, we fix the classical part of the scalar fluctuation of the metric around the string background configurations, which are parametrized by the superstring backgrounds, $G_{\mu\nu}(x)$ and $B_{\mu\nu}(x)$. We showed that the two-point correlation functions of the quantum parts of the scalar fluctuation are path-integrals of the  perturbative superstrings  on the string backgrounds. In this derivation, we move from the second quantization formalism to the first one, where  the coordinates of the two fields in the correlation functions become the asymptotic fields that represent the initial state $\bold{X}^{\mu}(\tau=-\infty, \sigma, \theta)$ and the final state $\bold{X}^{\mu}(\tau=\infty, \sigma, \theta)$, respectively.  All the paths on the string manifolds from $\bold{X}^{\mu}(\tau=-\infty, \sigma, \theta)$ to  $\bold{X}^{\mu}(\tau=\infty, \sigma, \theta)$ are summed up in the first quantization representation of the two-point correlation functions. Because the paths on the string manifolds are world-sheets with genera as shown in the section two in \cite{Sato:2017qhj}, they reproduce the  path integrals of the  perturbative strings up to any order, although the correlation functions are at tree level. 

Next task is 
to derive path-integrals of the perturbative heterotic strings  on all the string backgrounds, $G_{\mu\nu}(x)$, $B_{\mu\nu}(x)$, and $A_{\mu} (x)$
from the string geometry theory by considering the heterotic string manifolds.

\section*{Acknowledgements}
We would like to thank  
H. Kawai,
K. Kikuchi,
T. Yoneya,
and especially 
A. Tsuchiya
for discussions.
The work of M. S. is supported by Hirosaki University Priority Research Grant for Future Innovation.
The work of K. U. is supported by Grants-in-Aid from the Scientific 
Research Fund of the Japan Society for the Promotion of 
Science, under Contract No. 16K05364 and by the Grant ``Fujyukai'' 
from Iwanami Shoten, Publishers.

\appendix

\section{Green function on string geometry}
\setcounter{equation}{0}
In this appendix, we will show that  (\ref{GreenFunc}) is indeed a Green function on the flat superstring manifold. 
If ${\bf X}_{\hat{D}_{T}}^\mu(\bar{\sigma}, \bar{\theta})
\not\equiv {{\bf X}'}_{\hat{D}_{T}}^\mu(\bar{\sigma}, \bar{\theta})$\,,
we have 
\Eqr{
&&\hspace{-1cm}\frac{1}{\bar{e}'}
\frac{\pd}{\pd {\bf X}_{\hat{D}_{T}\,\nu}(\bar{\sigma}', \bar{\theta}')}
\,{\cal N}\,\left[\int d\bar{\sigma}d\bar{\theta}\,
\frac{\bar{e}^2}{\bar{\bf E}}
\left({\bf X}_{\hat{D}_{T}}^\mu(\bar{\sigma}, \bar{\theta})
-{{\bf X}'}_{\hat{D}_{T}}^\mu(\bar{\sigma}, \bar{\theta})\right)^2
\right]^{\frac{2-D}{2}}\nn\\
&&\hspace{-0.5cm}
=(2-D)\,{\cal N}\,\left[\int d\bar{\sigma}d\bar{\theta}
\frac{\bar{e}^2}{\bar{\bf E}}
\left({\bf X}_{\hat{D}_{T}}^\mu(\bar{\sigma}, \bar{\theta})
-{{\bf X}'}_{\hat{D}_{T}}^\mu(\bar{\sigma}, \bar{\theta})\right)^2
\right]^{-\frac{D}{2}}
\frac{\bar{e}'}{\bar{\bf E}'}
\left({\bf X}_{\hat{D}_{T}}^\nu(\bar{\sigma}', \bar{\theta}')
-{{\bf X}'}_{\hat{D}_{T}}^\nu(\bar{\sigma}', \bar{\theta}')\right)\,,
\nn\\
}
and then, 
\Eqr{
&&\frac{1}{\bar{e}''}
\frac{\pd}{\pd {\bf X}_{\hat{D}_{T}}^\nu(\bar{\sigma}'', \bar{\theta}'')}
\frac{1}{\bar{e}'}
\frac{\pd}{\pd {\bf X}_{\hat{D}_{T}\,\nu}(\bar{\sigma}', \bar{\theta}')}
\,{\cal N}\,\left[\int d\bar{\sigma}\,d\bar{\theta}
\frac{\bar{e}^2}{\bar{\bf E}}
\left({\bf X}_{\hat{D}_{T}}^\mu(\bar{\sigma}, \bar{\theta})
-{{\bf X}'}_{\hat{D}_{T}}^\mu(\bar{\sigma}, \bar{\theta})\right)^2
\right]^{\frac{2-D}{2}}\nn\\
&&~~~
=d(2-D)\frac{1}{\bar{\bf E}'}\frac{\bar{e}'}{\bar{e}''}\,{\cal N}\,
\left[\int d\bar{\sigma}\,d\bar{\theta}
\frac{\bar{e}^2}{\bar{\bf E}}
\left({\bf X}_{\hat{D}_{T}}^\mu(\bar{\sigma}, \bar{\theta})
-{{\bf X}'}_{\hat{D}_{T}}^\mu(\bar{\sigma}, \bar{\theta})\right)^2
\right]^{-\frac{D}{2}}
\delta(\bar{\sigma}'-\bar{\sigma}'')\,
\delta(\bar{\theta}'-\bar{\theta}'')
\nn\\
&&~~~~~
-D(2-D)\frac{1}{\bar{\bf E}'}\frac{1}{\bar{\bf E}''}\bar{e}'\bar{e}''
\,{\cal N}\,\left[\int d\bar{\sigma}\,d\bar{\theta}
\frac{\bar{e}^2}{\bar{\bf E}}
\left({\bf X}_{\hat{D}_{T}}^\mu(\bar{\sigma}, \bar{\theta})
-{{\bf X}'}_{\hat{D}_{T}}^\mu(\bar{\sigma}, \bar{\theta})\right)^2
\right]^{-\frac{D+2}{2}}\nn\\
&&~~~~~\times
\left({\bf X}_{\hat{D}_{T}}^\nu(\bar{\sigma}', \bar{\theta}')
-{{\bf X}'}_{\hat{D}_{T}}^\nu(\bar{\sigma}', \bar{\theta}')\right)
\left({\bf X}_{\hat{D}_{T}\,\nu}(\bar{\sigma}'', \bar{\theta}'')
-{{\bf X}'}_{\hat{D}_{T}\,\nu}(\bar{\sigma}'', \bar{\theta}'')\right)\,.
}
Thus, 
\Eqr{
&&\int d\bar{\sigma}'\,d\bar{\theta}'\,\bar{\bf E}'\,
\frac{1}{\bar{e}'}
\frac{\pd}{\pd {\bf X}_{\hat{D}_{T}}^\nu(\bar{\sigma}', \bar{\theta}')}
\frac{1}{\bar{e}'}
\frac{\pd}{\pd {\bf X}_{\hat{D}_{T}\,\nu}(\bar{\sigma}', \bar{\theta}')}
\,{\cal N}\,
\left[\int d\bar{\sigma}d\bar{\theta}
\frac{\bar{e}^2}{\bar{\bf E}}
\left({\bf X}_{\hat{D}_{T}}^\mu(\bar{\sigma}, \bar{\theta})
-{{\bf X}'}_{\hat{D}_{T}}^\mu(\bar{\sigma}, \bar{\theta})\right)^2
\right]^{\frac{2-D}{2}}\nn\\
&&
=d \int d\bar{\sigma}'\,
d\bar{\theta}'\,\mb{\delta}(0)\,(2-D)\,{\cal N}\,
\left[\int d\bar{\sigma}d\bar{\theta}
\frac{\bar{e}^2}{\bar{\bf E}}
\left({\bf X}_{\hat{D}_{T}}^\mu(\bar{\sigma}, \bar{\theta})
-{{\bf X}'}_{\hat{D}_{T}}^\mu(\bar{\sigma}, \bar{\theta})\right)^2
\right]^{-\frac{D}{2}}\nn\\
&&~~~
-D(2-D)\,{\cal N}\,
\left[\int d\bar{\sigma}d\bar{\theta}
\frac{\bar{e}^2}{\bar{\bf E}}
\left({\bf X}_{\hat{D}_{T}}^\mu(\bar{\sigma}, \bar{\theta})
-{{\bf X}'}_{\hat{D}_{T}}^\mu(\bar{\sigma}, \bar{\theta})\right)^2
\right]^{-\frac{D+2}{2}}\nn\\
&&~~~~~\times
\int d\bar{\sigma}'d\bar{\theta}'\,
\frac{\bar{e}'^2}{\bar{\bf E}'}
\left({\bf X}_{\hat{D}_{T}}^\nu(\bar{\sigma}', \bar{\theta}')
-{{\bf X}'}_{\hat{D}_{T}}^\nu(\bar{\sigma}', \bar{\theta}')\right)^2\nn\\
&&
=0\,,
}
where we use $D=d \int d\bar{\sigma}'\,
d\bar{\theta}'\,\mb{\delta}(0)$\,. 
Hence, we find
\Eqr{
&&\int d\bar{\sigma}'\,d\bar{\theta}'
\,\bar{\bf E}'\,\frac{1}{\bar{e}'}
\frac{\pd}{\pd {\bf X}_{\hat{D}_{T}}^\nu(\bar{\sigma}', \bar{\theta}')}
\frac{1}{\bar{e}'}
\frac{\pd}{\pd {\bf X}_{\hat{D}_{T}\,\nu}(\bar{\sigma}', \bar{\theta}')}
\mathcal{N}
\left[\int d\bar{\sigma}d\bar{\theta}\,
\frac{\bar{e}^2}{\bar{\bf E}}
\left({\bf X}_{\hat{D}_{T}}^\mu(\bar{\sigma}, \bar{\theta})
-{{\bf X}'}_{\hat{D}_{T}}^\mu(\bar{\sigma}, \bar{\theta})\right)^2
\right]^{\frac{2-D}{2}}\nn\\
&&~~~
=\delta({\bf X}_{\hat{D}_{T}}-{\bf X}_{\hat{D}_{T}}')\,,
}
where $\mathcal{N}$ is a normalizing constant.

\end{document}